\documentclass[11pt,a4paper]{article}
\usepackage[T1]{fontenc}
\usepackage{mathtools,amssymb,amsthm}
\usepackage[left=1in,right=1in]{geometry}
\usepackage{graphicx}
\usepackage{booktabs}
\usepackage[english]{babel}
\usepackage{xcolor,hyperref}

\hypersetup{colorlinks, linkcolor=blue}

\renewcommand*{\th}{\textsuperscript{th}}
\newcommand*{\R}{\mathbb R}
\newcommand*{\C}{\mathcal C}

\newcommand*{\phv}{\,\cdot\,} 

\newcommand*{\pos}{+}
\renewcommand{\neg}{-}
\newcommand*{\lampos}{\lambda^\pos}
\newcommand*{\lamneg}{\lambda^\neg}
\newcommand*{\gpj}[1][J]{g_{#1}}
\newcommand*{\gnj}[1][J]{h_{#1}}
\makeatletter
\newcommand*{\wpos}{\@ifstar\@@wpos\@wpos}
\newcommand*{\@wpos}{w_\pos}
\newcommand*{\@@wpos}{{\@wpos}_*}
\newcommand*{\wneg}{\@ifstar\@@wneg\@wneg}
\newcommand*{\@wneg}{w_\neg}
\newcommand*{\@@wneg}{{\@wneg}_*}
\newcommand*{\w}{\@ifstar\@@w\@w}
\newcommand*{\@w}{\mathbf{w}}
\newcommand*{\@@w}{\mathbf{w}_*}
\newcommand*{\dw}{\@ifstar\@@dw\@dw}
\newcommand*{\@dw}{d\mathbf{w}}
\newcommand*{\@@dw}{d\mathbf{w}_*}
\makeatother
\newcommand*{\W}{\mathbf W}
\newcommand*{\der}[2]{\frac{d #1}{d #2}}
\newcommand*{\pd}[2]{\frac{\partial#1}{\partial #2}}
\newcommand{\intz}{\int\limits_{\mathclap{{[0,1]}}}}
\newcommand{\intzz}{\int\limits_{\mathclap{{[0,1]^2}}}}
\renewcommand*{\(}{\begin{equation}}
\renewcommand*{\)}{\end{equation}}
\let\bar\overline

\allowdisplaybreaks[4]

\DeclarePairedDelimiter{\abs}{\lvert}{\rvert}
\DeclarePairedDelimiter{\set}{\lbrace}{\rbrace}

\title{Modeling opinion polarization on social media: 
application to Covid-19 vaccination hesitancy in Italy}

\author{Jonathan Franceschi\thanks{Department of Mathematics \lq\lq F. 
Casorati\rq\rq, University of Pavia, Italy\newline\leavevmode\hspace*{1.5em} 
({\ttfamily jonathan.franceschi01@universitadipavia.it})}
\and Lorenzo Pareschi\thanks{Department of Mathematics and Computer Science, 
University of Ferrara, Italy\newline\leavevmode\hspace*{1.5em}
({\ttfamily lorenzo.pareschi@unife.it})}
\and Elena Bellodi\thanks{Department of Engineering, University of Ferrara, Italy ({\ttfamily elena.bellodi@unife.it})}
\and Marco Gavanelli\thanks{Department of Engineering, University of Ferrara, Italy ({\ttfamily marco.gavanelli@unife.it})} 
\and Marco Bresadola\thanks{Department of Humanities, University of Ferrara, 
Italy ({\ttfamily marco.bresadola@unife.it})}
}
\date{}

\begin{document}
\maketitle
\begin{abstract}
The SARS-CoV-2 pandemic reminded us how vaccination can be a divisive 
topic on which the public conversation is permeated by misleading claims, 
and thoughts tend to polarize, especially on online social networks. In this 
work, motivated by recent natural language processing techniques to systematically 
extract and quantify opinions from text messages, 
we present a differential framework for bivariate opinion formation 
dynamics that is coupled with a compartmental model for fake news 
dissemination. Thanks to a mean-field analysis we demonstrate that the resulting
Fokker-Planck system permits to reproduce 
bimodal distributions of opinions as observed in polarization dynamics.
The model is then applied to sentiment analysis data from  
social media platforms in Italy, in order to analyze the evolution of opinions 
about Covid-19 vaccination. We show through numerical simulations that the model is capable to describe 
correctly the formation of the bimodal opinion structure observed in the 
vaccine-hesitant dataset, which is witness of the known 
polarization effects that happen within closed online communities.
\end{abstract}


\tableofcontents

\section{Introduction}

Vaccination coverage, globally, has been at its highest levels for the last 
decades, with the notable exception of measles and diphtheria^^>\cite{who22}. 
Measles outbreaks in particular^^>\cite{cdc22} have raised concern in the 
Western public since they were about a disease that vaccination and treatments 
had reduced to a condition of rarity.
The SARS-CoV-2 pandemic has been one compelling reason to rethink vaccination 
as an effective medical practice to prevent the spreading of diseases, 
especially in relation to the massive media coverage of the topic. The issue, 
in this case, is that a polarizing debate could exacerbate \emph{vaccination 
hesitancy}, i.e., the reluctance in getting 
vaccinated (see^^>\cite{ren18,roy22,muller22,salathe08}, but also the 
recent^^>\cite{dror20,schmidt18,cascini22}), with potentially dangerous 
implications for healthcare^^>\cite{alvarez17,funk10,yaqub14,troiano21}. 
Moreover, unstable contexts like this one are more likely to develop 
irreparable fractures when misinformation 
is disseminated among people, and a positive reinforcement loop clusters the 
audience into isolated groups (the so-called \emph{echo chambers}) where the 
only information shared is the one aligned with the majority point of 
view^^>\cite{tornberg21,bessi16,muller22,cinelli21}.  
Therefore, the need for the policymaker to be able to take informed decisions 
based on the understanding of the directions of the debate evolution is 
critical.  

Vaccines are no stranger issue to mathematical literature, 
too^^>\cite{fu11,fine86}, 
especially the game-theoretical one. The \emph{free-ride effect} of the portion 
of population not taking the possible risks associated with vaccination to 
enjoy the benefits given by the herd immunity are well 
known^^>\cite{galvani07,bauch03}. So are 
opinions^^>\cite{pires18,sznajd05,axelrod97,garnier17,noorazar20}, especially 
in the 
context of multi-agent systems and kinetic 
theory more 
generally^^>\cite{albi15,pareschi17,mastroeni19,goddard22,toscani06,pareschi13}. Within the 
same field, works have been proposed 
recently that borrow from the classical compartmental framework of 
epidemiological theory^^>\cite{heth00,daley64}, both for the spread of diseases and 
misinformation^^>\cite{dimarco20,fp22,fpz22,zan22,bern22}.

In this paper, we build on these elements to present a differential model, 
based on a mean-field description of agent dynamics, for the evolution of 
opinions in the presence of fake news spreaders. Although the model is mainly 
applied to the spread of fake news on social networks regarding the hesitancy 
to Covid-19 vaccination, its structure easily finds application in more general 
contexts where polarization of opinions within closed online communities is observed.

More in details, following the seminal paper^^>\cite{fp22}, we consider a multi-agent population with a structure, 
where the modeling of fake news dissemination is managed via a set of 
compartments. In this setting, where the fake-news is treated as the spreading of
a virus, the underlying variable that is shared by each agent 
characterizes a bivariate opinion distribution, that takes into account both positive and negative 
opinions about a given topic. This last aspect is crucial with a view to aligning the model with 
experimental data from \emph{sentiment analysis} carried out on social media platforms, such as blogs
and social networks. In such a situation, each opinion is inherently two-dimensional, as it  
classifies the polarity of a given text according to which level the opinion expressed is positive or negative^^>\cite{yue19,nemes21,melton22,piedrahita21,nicola18}.

We emphasize that, unlike^^>\cite{albi17,fp22,toscani06}, our starting point is a 
system of stochastic differential equations (SDEs) for the dynamics of opinions 
and not a binary interaction dynamics leading to a Boltzmann-type equation in 
the limit of a large number of agents. In fact, we are interested in modeling a 
situation in which agents interact simultaneously with the entire population a 
scenario typical of group chats in instant messaging.
As a consequence, the resulting model 
can be analyzed directly thanks to its mean-field approximation that permits
to compute explicitly the steady states of the system without resorting to the quasi-invariant opinion approximation. The equilibrium states, in contrast with the classical case \cite{sznajd00, sznajd05, toscani06}, are characterized by
a superposition of Beta distributions that give rise to bimodal shapes,
i.e., individuals' thoughts polarizing around different extreme positions, with 
a certain absence of compromise, in agreement with those observed opinion 
polarization effects in closed communities.

The model is then interfaced with available data concerning Covid-19 vaccination 
in Italy from the popular messaging app  Telegram; one of its features is the possibility of having large online group 
chats focused on a topic of choice. They effectively form closed communities, 
where conversations experiment a low degree of noise: they are therefore ideal 
to analyze the evolution of sentiments about a certain subject. Numerical simulations show the model's ability to interface correctly with the data extracted using NLP techniques and to describe the polarization phenomenon over time very well. 

The rest of the manuscript is organized as follows: 
in Section^^>\ref{sec:stocmodel}, we present the stochastic differential 
model for opinion-formation processes characterized by two-dimensional vectors, 
which in the mean-field limit is approximated by a Fokker-Planck equation that 
allows us to compute steady states for marginals in explicitly solvable special cases.
Next, in Section^^>\ref{sec:opinionfake}, we merge the model 
with a compartmental framework to take into account the 
potential spread of misinformation which can act as a catalyst for the 
polarization. In Section^^>\ref{sec:data} we present the social media dataset 
and compare the evolution predicted by the model to the data one. Finally, in 
the last section some final considerations and concluding remarks are reported.

\section{Mean-field models of bivariate opinion formation}\label{sec:stocmodel}

When modelling the dynamics of opinions within individuals from a mathematical point of view, several approaches based on multi-agent interactions at various levels are possible^^>\cite{deffuant00,albi15,sznajd00,deffuant00,tornberg21,albi17,aletti07}. It 
is customary to set the interval $[-1, 1] \subseteq \R $ as a natural space for 
the variable $w$ representing the opinion, intending that radical positions are 
assumed as the absolute value $\abs w$ approaches^^>$1$, while neutral ones are 
assumed near^^>$0$. This choice embeds opinions as a continuous spectrum 
between positive and negative convictions and allows for a relatively simple 
description as a one-dimensional variable.

Here, we are setting ourselves in a subspace of the plane to better interpret 
the inherently two-dimensional nature of the description of opinions given by 
\emph{natural language processing} (NLP) techniques like sentiment analysis, which assign scores based on how much a 
certain thought can be perceived as positive or negative, so that each record 
is associated with a pair of scores. Although a multivariate model for opinion 
requires greater care to devise it and to perform computations, it also gives 
us more coherent informations when aligning the model with data. One possibility would have been to simply map our 
into $[-1,1]$, but this would inherently involve mapping each individual's pair 
of sentiments into two opposing opinions that would statistically correspond to 
two different individuals.

\begin{figure}
\centering
\includegraphics[width=0.99\linewidth]{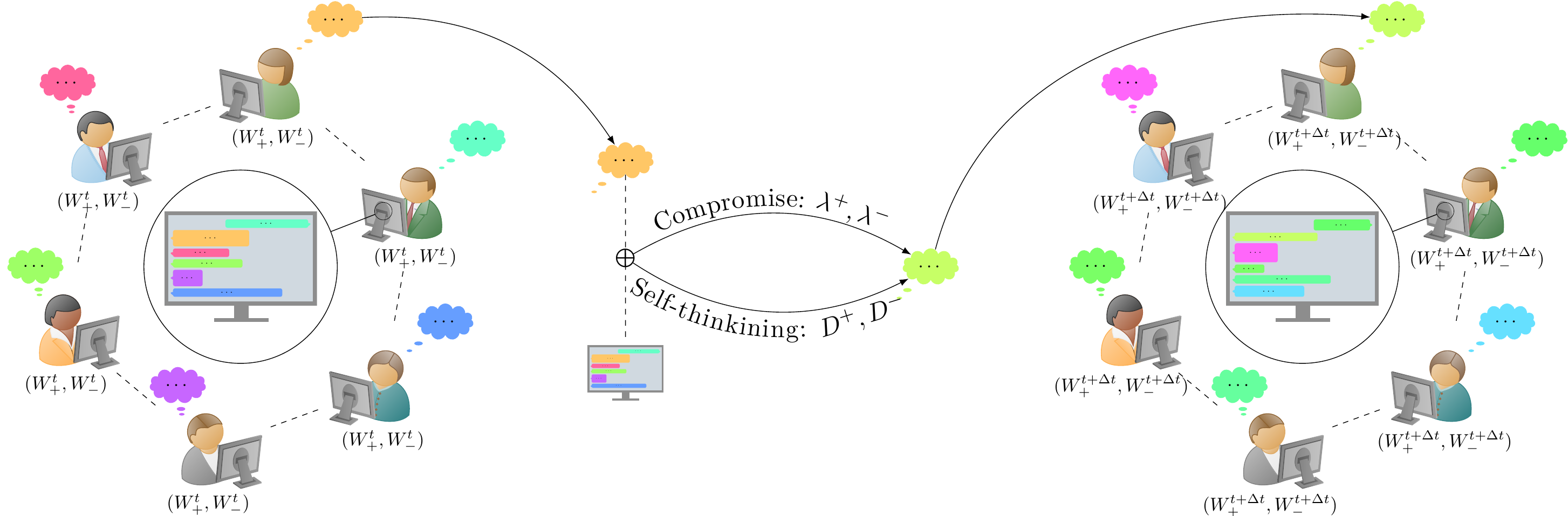}
\caption{Opinions expressed by users in a group chat (represented here by 
different colors) are extrapolated by the messages via NLP and are denoted by a pair of continuous, 
time-dependent, real values $(W_t^\pos,W_t^\neg)\in [0,1]^2$: one for how positive the opinion is and one 
for how negative it is, respectively. The dynamics is then characterized by the functions $\lambda^\pos,\lambda^\neg$ that define the compromise process and 
the functions $D^\pos,D^\neg$ that formalize individual self-thinking.}
\end{figure}

The other main modeling choice is the use of SDEs instead of other 
alternatives, such as binary interactions described according to a broader 
interpretation of particle dynamics typical of statistical mechanics. Here we 
are interested in modeling a situation in which agents interact simultaneously 
with the whole population at all times: this is the typical scenario of group 
chats within instant messaging applications.

\subsection{A multi-agent stochastic differential model}
We consider a population of $N$ indistinguishable agents characterized by the opinion vector  
variable $\w = (\wpos,\wneg) \in [0,1]^2$ that represents both positive and 
negative opinions of each agent. The closer you get to $0$ the milder the opinion, and the closer you get to 
$1$ the more extreme. If we indicate with 
${\W}^i_t=(W_{\pos,t}^i,W_{\neg,t}^i)$ the pair of opinions associated 
with the $i$-th agent at the instant $t\ge 0$, its continuous time evolution 
can be expressed via a stochastic differential system of the general form
\(
\left\lbrace
\begin{aligned}
 d W_{\pos,t}^i &= \frac{1}{N}\sum_{j=1}^N \Lambda^+({\W}_{t}^i,{\W}_{t}^j) 
 (W_{\pos,t}^j - W_{\pos,t}^i)+ \sigma_\pos D^\pos({\W}_{t}^i) dB_{\pos,t}^i\\
 d W_{\neg,t}^i &= \frac{1}{N}\sum_{j=1}^N \Lambda^-({\W}_{t}^i,{\W}_{t}^j) 
 (W_{\neg,t}^j-W_{\neg,t}^i)+ \sigma_\neg D^\neg({\W}_{t}^i) dB_{\neg,t}^i
\label{eq:sde}
\end{aligned}
\right.
\)
where $\Lambda^{\pm}({\W}_{t}^i,{\W}_{t}^j)$ are nonnegative functions 
characterizing  
the rate towards compromise when two agents $i$ and $j$ interact. 
Then $\sigma_{\pm}$ are 
positive constant diffusion coefficients, while the nonnegative functions 
$D^\pm({\W}_{t}^i)$ represent the local incidence of the diffusion effects due 
to self-thinking of agent $i$.  The latter functions usually vanish at the boundary 
of $[0,1]^2$ so that people at extreme positions are less subject to noise effects. 
Finally, $dB_{\pm,t}^i$ are independent one-dimensional Brownian 
motions to take into account the random nature of social interactions.
It should be noted that due to the presence of noise, the dynamics in \eqref{eq:sde} may give rise to an inadmissible opinion vector. For this reason, the model is supplemented with appropriate boundary conditions that constrain the opinion vector within the domain $[0,1]^2$. 

A typical example of compromise functions is represented by the bounded confidence model^^>\cite{hegselmann02} where agents interact only if their opinions differ no more than a certain confidence level $\Delta_\pm \in [0,1]$   
\[
\Lambda^{\pm}({\W}_{t}^i,{\W}_{t}^j) = \lambda^\pm \Psi(|{W}_{\pm,t}^i-{W}_{\pm,t}^j|\leq \Delta_\pm)
\]
with $\Psi(\cdot)$ the indicator function and $\lambda^\pm$ suitable positive constants.
In the sequel, we will restrict to the simplified situation where $\Delta_\pm = 1$ and thus $\Lambda^{\pm}({\W}_{t}^i,{\W}_{t}^j)=\lambda^\pm$ represent the alignment strengths towards the current 
mean positive and 
negative opinions of the population
\begin{equation} 
M^N_{\pos,t}=\frac1{N} \sum_{j=1}^N W_{t,\pos}^j,\qquad M^N_{\neg,t}=\frac1{N} 
\sum_{j=1}^N W_{t,\neg}^j.
\label{eq:Mpn}
\end{equation}
Next, we introduce the empirical measure 
\(
f^N(\w,t) \coloneqq \frac1N \sum_{i=1}^N 
\delta(\w-{\W}_t^i),
\) 
where $\delta(\cdot)$ is the Dirac delta function, which counts how many agents share the same pair of opinions 
at time $t\ge0$. Our main goal is to analyze the evolution of the empirical 
measure of the system, especially when the number of individuals in the 
population grows large. The advantage of resorting to a limit procedure in order to study the 
mean-field version of system^^>\eqref{eq:sde} is that under simplifying 
hypotheses it is possible to compute the stationary state of quantities of 
interest concerning the system. 

\begin{figure}[t]
\centering
\includegraphics[width=\linewidth]{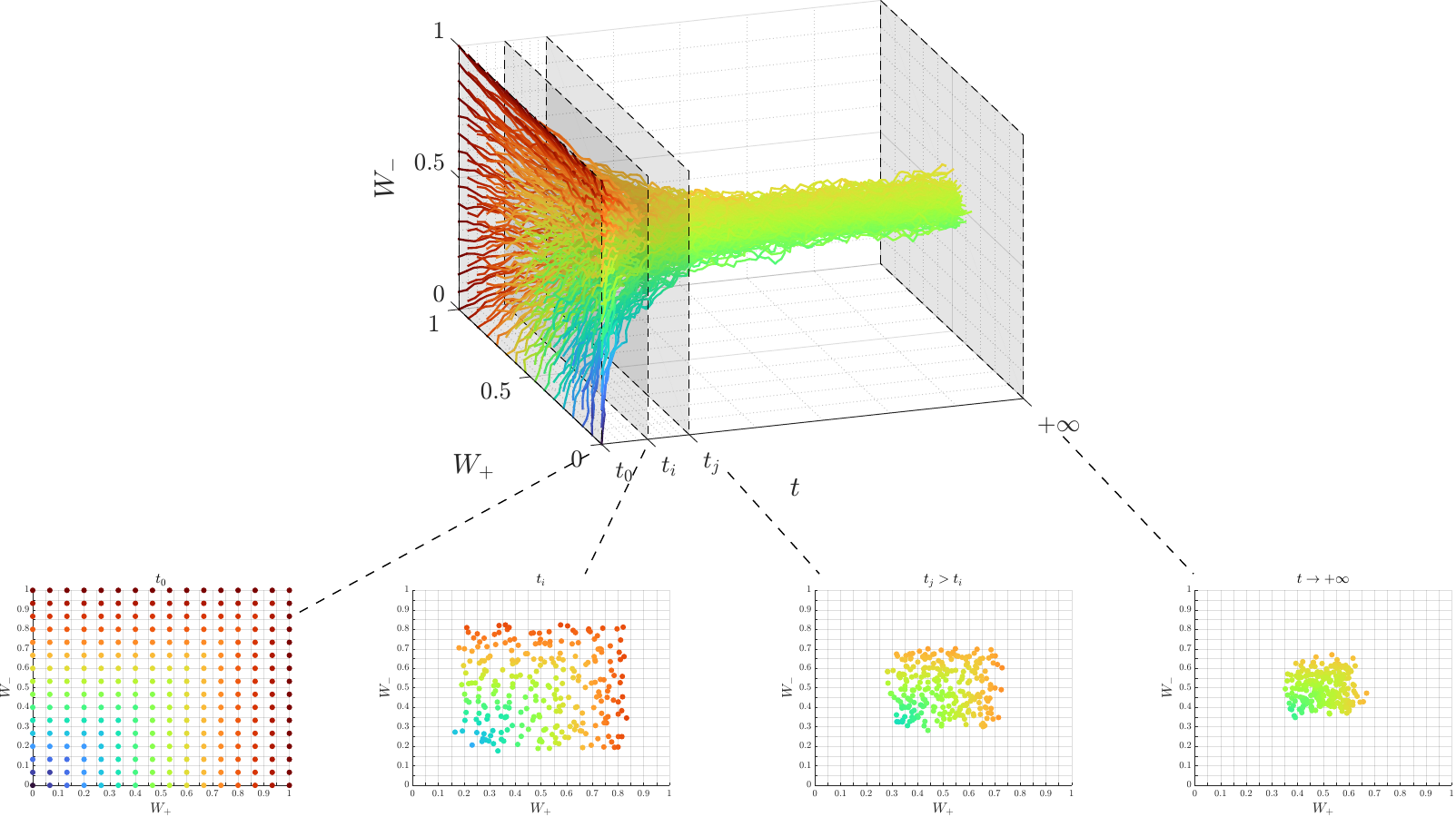}
\caption{Simulation of model^^>\eqref{eq:sde} using Euler-Maruyama scheme using $N=256$ agents with
$\lampos=\lamneg=\sigma_+=\sigma_-=0.05$. Here colors 
represent 
opinion's intensity measured as $\Vert{\mathbf W_t}\Vert_\infty$. Opinions  
initially are drawn from a uniform distribution and then, due to the compromise
dynamics, concentrate toward the center of the unit square.}
\end{figure}

\subsection{Mean-field limit and Fokker-Planck approximation}

A classical approach to formally analyze the behavior of the system when the number $N$ of agents in the population is large,
is to consider the $N$-particle probability density^^>\cite{golse,Jabin2017}
\[
f^{(N)}({\W}_t^1,\ldots, {\W}_t^N,t)
\]
and the associated first marginal 
\[
f_1^{(N)}({\W}_t^1,t)=\int_{[0,1]^{2N-2}} f^{(N)}({\W}_t^1,\ldots, 
{\W}_t^N,t)\,d{\W}^2_t,\ldots,d{\W}_t^N.
\]
and make the so-called \emph{propagation of chaos assumption} on the marginals.
More specifically, we assume that $f^{(N)} \approx f^{\otimes N}$ for $N \gg 
1$, i.e., the random vectors ${\W}_t^1,\ldots, {\W}_t^N$ are approximately 
independently $f(\w,t)$-distributed.

In this case, for $N \gg 1$ we can write
\(
f^N(\w,t) \approx  f(\w,t),\qquad (M^N_{\pos,t},M^N_{\neg,t}) \approx 
(m^\pos (t),m^\neg (t)) = \int_{[0,1]^2} f(\w,t) \w\,d\w,
\)
due to the law of large numbers. Consequently, the SDE model \eqref{eq:sde} becomes
independent of $j\neq i$ and we obtain the so-called \emph{Mc-Kean nonlinear process} which, in the simplified situation where $\lambda^\pm$ are non negative constants, reads
\(
\left\lbrace
\begin{aligned}
 dW^\pos_t &= \lampos (m^\pos (t)-W^\pos_{t})+ \sigma_\pos 
 D^\pos({\W}_{t}) dB^\pos_{t},\\
 dW^\neg_t &= \lamneg (m^\neg (t)-W^\neg_{t})+ \sigma_\neg 
 D^\neg({\W}_{t}) dB^\neg_{t},
 \label{eq:sdeMK}
\end{aligned}
\right.
\)
with $f={\rm law}({\W}_t)$.
The above system may be equivalently expressed by  
a nonlinear Fokker-Planck equation of the form^^>\cite{risken96,golse,Jabin2017} 
\(\label{eq:FPopinion}
\begin{aligned}
\pd{}{t} f(\w,t) &= \lampos\pd{}{\wpos}[\bigl(\wpos - 
m^\pos (t)\bigr) f(\w,t)] + \lamneg\pd{}{\wneg}[\bigl(\wneg - 
m^\neg (t)\bigr) f(\w,t)] \\
&\hphantom{{}=} + \frac{\sigma_\pos^2}{2} 
\pd{^2}{\wpos^2}\bigl(D^+(\w)^2 f(\w,t)\bigr) + 
\frac{\sigma_\neg^2}{2}
\pd{^2}{\wneg^2}\bigl(D^-(\w)^2 f(\w,t)\bigr).
\end{aligned}
\)
Equation^^>\eqref{eq:FPopinion} needs to be complemented with suitable no-flux boundary conditions that guarantee
$f(\w,t)$ to be compactly supported in $[0,1]^2$
\(\label{eq:boundcond}
\begin{aligned}
\lampos\bigl(\wpos - m^\pos (t)\bigr) f(\w,t)
+ \frac{\sigma_+^2}{2} \pd{}{\wpos} \bigl(D^\pos(\w)^2 f(\w,t)\bigr) = 0,\quad {{\rm on\,\,} w_+=0,1}\\
\lamneg\bigl(\wneg - m^\neg (t)\bigr) f(\w,t)
+ \frac{\sigma_-^2}{2} \pd{}{\wneg} \bigl(D^\neg(\w)^2 f(\w,t)\bigr)=0 ,\quad {{\rm on\,\,} w_-=0,1}.
\end{aligned}
\)
Thanks to the above conditions we can introduce the normalization assumption
\[
\int_{[0,1]^2} f(\w,t) \dw = 1, \quad\forall\, t \ge 0.
\]
We refer to \cite{Jabin2017}, and the references therein, for rigorous results
concerning the mean-field limit of stochastic particle system of type \eqref{eq:sde}.

Note that, if in addition, at the boundary of $[0,1]^2$ we have 
\(
D^+(\w)^2f(\w,t)=D^-(\w)^2f(\w,t)=0,
\label{eq:bd2}
\) 
integrating by parts and using the no-flux boundary conditions 
\eqref{eq:boundcond}, we have conservation of the mean opinion  
\[
\frac{d\, {\bf m} (t)}{dt} = \int_{[0,1]^2} \frac{\partial}{\partial t}f(\w,t)\w\,d\w=0.
\]
If we now define the variances of the variables $w^+$ and $w^-$ as
\[
V^+ (t) = \int_{[0,1]^2} f(\w,t)(w^+-m^+ )^2\,d\w,\quad V^- (t) = \int_{[0,1]^2} f(\w,t)(w^--m^- )^2\,d\w,
\]
we have
\(
\frac{d V^\pm (t)}{dt} = -2\lambda^\pm V^\pm (t) + \sigma_\pm^2 \int_{[0,1]^2} D^+(\w)^2 f(\w,t)\,d\w.
\label{eq:vdc}
\)
This shows that the particular choice of the functions $D^\pm(\w)$ influences the behavior of the variance and so the 
convergence to equilibrium of the Fokker-Planck equation \eqref{eq:FPopinion}.

\subsection{Equilibrium states for the marginal densities}
The functions $D^\pm(\w)$ characterizing the local effect of diffusion, and 
thus the individual behavior of agents, 
turn out to be essential for the purpose of studying the equilibrium states of the system. For example taking
\[
D^\pm=|w^\pm-m ^\pm|,
\]
where agents tends to reduce self-thinking as their opinion is close to the average, from \eqref{eq:vdc} we get the uniform decay of the  variances as soon as $2\lambda^\pm > \sigma_\pm^2$. In this case the long time behavior is characterized by a Dirac delta function $f^\infty(\w)=\delta(\w-{\bf m} )$ where all agents are concentrated on the same opinion. 

In the sequel, we assume that opinions close to zero and one are less prone to random opinion effects, 
in the sense that both very moderate and more extreme individuals in expressing opinions have less freedom to change opinion, since they are already positioned in an extremal state. This assumption turns out to be essential in order to derive steady states in agreement with the experimental data, and differs from classical one-dimensional opinion models where individuals with an opinion around zero are assumed to be hesitant and so mostly prone to the effect of diffusion.

To this aim, we consider the local diffusion function to be such that $D^\pos(\w) = D(\wpos)$ and $D^\neg(\w) = D(\wneg)$ with
\(
D(w)= \sqrt{w(1-w)},
\label{eq:diff}
\) 
so that it vanishes in $0$ and $1$. This assumption, if the solution $f(\w,t)$ is sufficiently regular, guarantees conditions \eqref{eq:bd2} at the boundary and therefore the mean opinion is independent from time.
We refer to^^>\cite{toscani06} for other admissible choices leading to interesting steady states. As we will see, thanks to \eqref{eq:diff} we are able to compute explicitly the steady state for the marginal densities. 

\begin{figure}\centering
\newsavebox\figuno
\savebox\figuno{\includegraphics[width=0.45\linewidth]{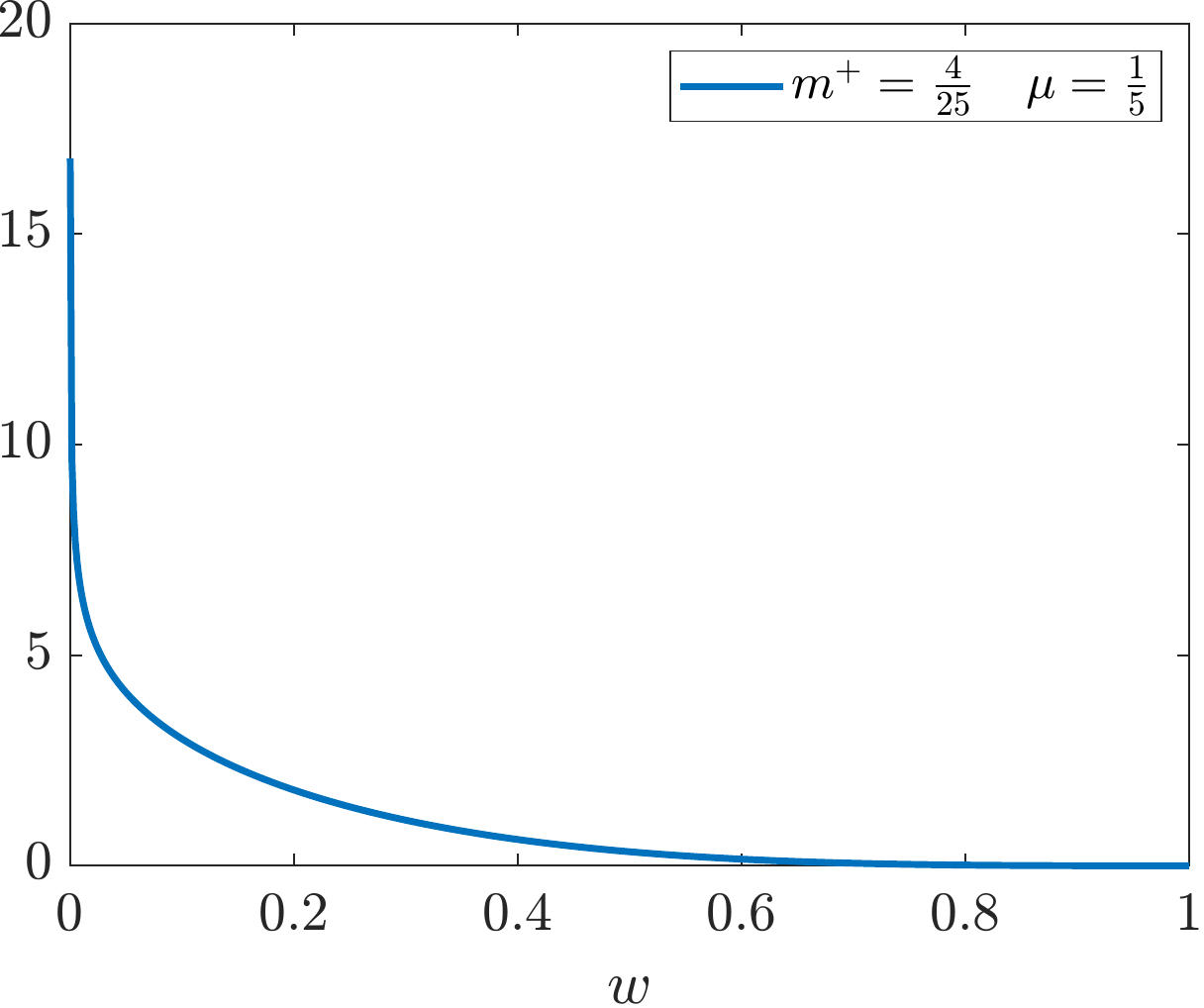}}
\usebox\figuno%
\hskip.5cm
\includegraphics[width=0.45\linewidth,height=\ht\figuno]{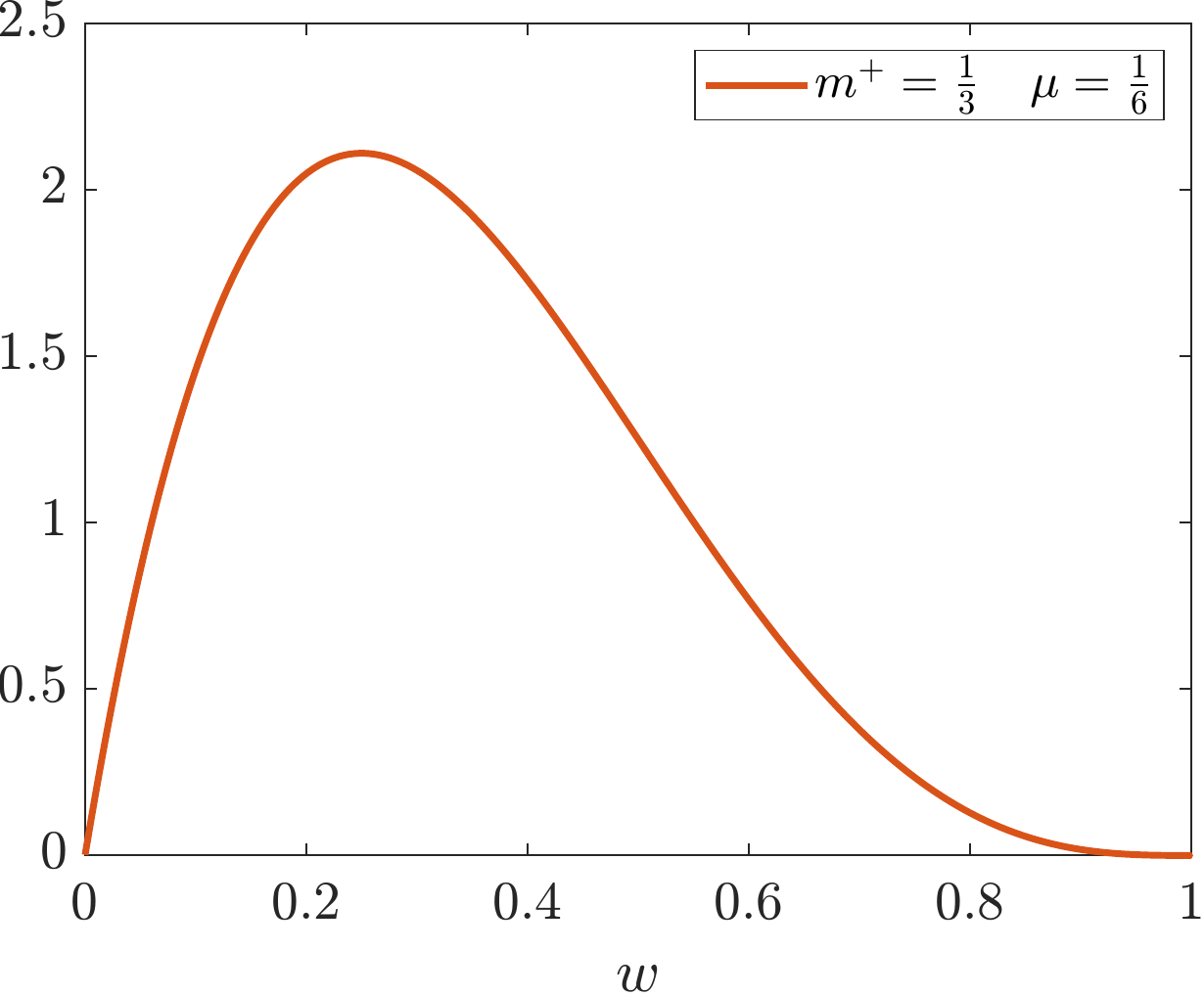}
\caption{Examples of stationary marginal opinion distributions $g^\infty(w_+)$ characterized by ${\rm Beta}(w_+; a,b)$ functions obtained with different 
choices of $m^+$ and $\mu$. Left: $m^+=4/25$, $\mu=1/5$, 
corresponding to $a=0.8$, $b=4.2$. We can see that the 
function tends to infinity as we approach the left boundary. Right: 
$m^+=1/3$, $\mu=1/6$, corresponding to $a=2$, $b=4$, with a unimodal 
structure. }
\label{fig:betasingle}
\end{figure}

Indeed, let us integrate directly system^^>\eqref{eq:FPopinion} with respect to 
the 
negative opinion $\wneg$, so that we have
\(
\begin{aligned}
\pd{}{t} \int_0^1 f(\w,t)\, d\wneg &= \lampos \pd{}{\wpos} 
\left[\bigl(\wpos - {m}^\pos\bigr)\int_0^1  f(\w,t)\, 
d\wneg\right]\\
&\hphantom{{}=} + \left.\lamneg\bigl(\wneg - 
m^\neg\bigr)f(\w,t)\right\vert_0^1\\
&\hphantom{{}=} + \frac{\sigma_\pos^2}{2}\pd{^2}{\wpos^2} \left[ 
\wpos(1 - \wpos) \int_0^1 
f(\w,t)\, d\wneg\right]\\
&\hphantom{{}=} + \left.\frac{\sigma_\neg^2}{2}
\pd{}{\wneg} \left[\wneg(1-\wneg) f(\w,t)\right]\right\vert_0^1\,.
\end{aligned}
\)
Thanks to the boundary conditions in^^>\eqref{eq:boundcond} we have the 
simplification
\(
\pd{}{t} g(\wpos,t) = \lampos \pd{}{\wpos} \left[
\bigl(\wpos - {m}^\pos\bigr) g(\wpos,t)\right] + 
\frac{\sigma_\pos^2}{2} \pd{^2}{\wpos^2} \left[
\wpos(1-\wpos) g(\wpos,t)\right],
\label{FPgpos}
\)
where we denote the marginal density of 
the positive opinion as 
\[
g(\wpos,t) = \int_0^1 f(\w,t) \, d\wneg.
\] 
We can now 
compute the stationary solution $g^\infty(\wpos)$ by observing that using the boundary conditions \eqref{eq:boundcond} it 
satisfies
\[
\lampos \left[
\bigl(\wpos - {m}^\pos\bigr) 
g^\infty(\wpos)\right] + 
\frac{\sigma_\pos^2}{2} \pd{}{\wpos} \left[
\wpos(1-\wpos) g^\infty(\wpos)\right] = 0.
\]
Thus, $g^\infty$ is computed explicitly as^^>\cite{toscani06} 
\(\label{eq:steadysingle}
g^\infty(\wpos)=  
C_\pos 
\wpos^{ {m}^\pos/\mu_\pos-1}
(1-\wpos)^{(1- {m}^\pos)/\mu_\pos-1},
\)
where $\mu_\pos \coloneqq \lampos/\sigma_\pos^2$, $m_{+}\in (0,1)$ and $C_\pos$ is a 
normalization constant which depends on all parameters appearing 
in^^>\eqref{eq:steadysingle}. Equation \eqref{eq:steadysingle} represents a Beta distribution
of the form ${\rm Beta}(w; a,b)$ with $a=m^+/\mu_+$ and 
$b=(1-m^+)/\mu_+$ (see Fig \ref{fig:betasingle}). Note also that since $a,b > 0$ condition \eqref{eq:bd2} is always guaranteed. 

Analogous 
calculations can be performed to obtain a closed expression for the steady 
state of the marginal density of the negative opinion^^>$h(\wneg,t)$ which reads
\(\label{eq:steadysingle2}
h^\infty(\wneg)=  
C_\neg 
\wneg^{ {m}^\neg/\mu_\neg-1}
(1-\wneg)^{(1- {m}^\neg)/\mu_\neg-1},
\)
where now $\mu_\neg \coloneqq \lamneg/\sigma_\neg^2$,  $m_{-}\in (0,1)$ and $C_\neg$ is a 
normalization constant. 

In Fig.^^>\ref{fig:betasingle} are reported examples of various stationary 
marginal opinion distributions. One of their interesting properties is that they are flexible 
enough to give rise to several different shapes, including some that are 
unbounded near the ends of the support. This is representative of extreme 
polarization phenomena in which the vast majority of the population shares 
extreme ideas. Beta distributions are also \emph{unimodal}, i.e., agents well 
described by a Beta tend to aggregate around a certain, unique, value, where 
this tendency depends on the variance of the distribution. However, this 
implies that a similar model would not be able to accurately describe different 
kinds of polarization, the ones that are local to certain population subsets
and that may have a \emph{bimodal} structure.

These latter phenomena are the ones more commonly associated with the formation 
of separate clusters within communities after the population has been exposed 
to fake news: misinformation exacerbates underlying radical opinions in a group 
of individuals, who then progressively proceed to discard any other belief. 
While this happens broadly, at different levels, (the phenomenon of \emph{echo 
chambers}), the less moderate ones are often the ones that raise more 
concern^^>\cite{chiou18,carrieri19,catalan20,waszak18,lyu21,montagni21}.

Therefore, the model should also take into account the effects brought by the 
dissemination of fake news that also lead to changes of the average opinions within subgroups
of individuals. This will be explored in the next section.

\section{Merging opinion formation with fake news dissemination}\label{sec:opinionfake}

The full model focuses again on a structured population where $N$ 
indistinguishable agents all share a vector-valued variable $\w \in [0,1]^2$. 
In the current setting, concurrent to the opinion formation process is also 
spread of misinformation, whose dynamics can be fruitfully approached through 
the compartmental framework typical of epidemiology^^>\cite{daley64,fp22,heth00}.

\subsection{Defining fake news}

Defining what fake news is and why it is a phenomenon deserving its own 
category (think for instance to other classifications of lies, e.g., scams, 
hoaxes, urban legends\ldots) is itself challenging. Our approach will be to 
consider fake news any piece of information whose \emph{initial} diffusion is 
made with the purpose of mislead people intentionally. The word \lq 
initial\rq\ here is key, because fake news is most often spread by people who 
do not know (or care) it is false. This has been linked to the concept of 
\emph{post-truth} and explored also from a philosophical point of 
view^^>\cite{waisbord18}; see^^>\cite{fp22} and the references therein for a 
recent overview on the different approaches for fake news detection.
\begin{figure}
\centering
\includegraphics[width=0.7\linewidth]{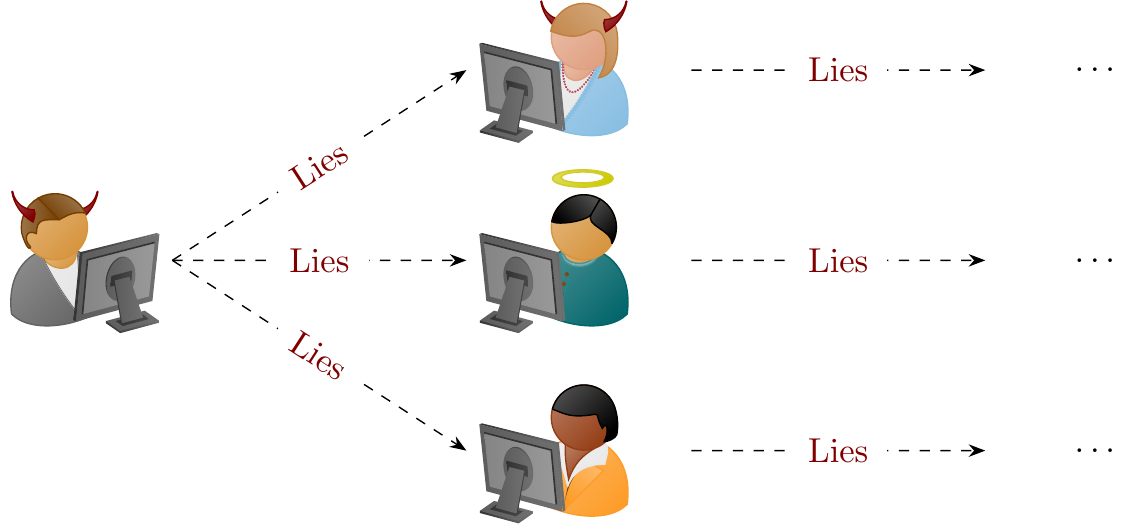}
\caption{Defining fake news: here the evil look of the person on the left 
symbolizes the purpose of voluntarily mislead others with false information; 
whereas the first recipients of the news spread it, in turn, either animated by 
the same will (evil agent, top), by the desire of sharing helpful or otherwise 
legitimate information (angelic agent, middle) or finally guided by no specific 
goal (neutral agent, bottom).}
\end{figure}

\subsection{A compartmental model for fake news spreading}
For what concerns the dissemination of fake news, we assume that the people 
within the community can be divided into four disjoint compartments: 
susceptible (or ignorant), exposed (or incubator), infected (or spreader) and recovered (or stifler). 
Concerning the nomenclature, we opted for the epidemiological convention here to adhere to the previous 
works^^>\cite{fp22,fpz22}. We refer to^^>\cite{fp22} and the references therein for other popular choices.

Then we use suitable differential 
equations to describe the way individuals change compartment. 
To each compartment will be 
assigned its initial as identifying letter, so that we will refer to them as 
the set^^>$\C\coloneqq \set{S, E, I, R}$. We shall therefore 
study the evolution of the opinion's distribution of the agents in each 
compartment, noted, respectively, by \mbox{$f_S = f_S(\w, t)$}, \mbox{$f_E = 
f_E(\w,t)$}, \mbox{$f_I = f_I(\w,t)$}, and \mbox{$f_R = f_R(\w,t)$}.

Like before, we restrict ourselves to consider a reduced time-span, during 
which we can assume that the population is fixed, i.e., nobody enters or leaves 
it; This choice is based on the average lifespan of fake news. Thus, we set the 
overall opinion distribution as a probability density for all $t \ge 0$, i.e.,  
\[
\intzz\ \sum_{J \in \C}f_J(\w,t)\, \dw = 1,\quad t > 0.
\]
The quantities in the first column of
\begin{align*}
\gpj(\wpos,t) &= \intz f_J(\w,t)\, d\wneg
& m_J^\pos(t)&= \frac1{\rho_J(t)}\ \intzz \wneg\, f_J(\w,t)\, \dw,\\
\gnj(\wneg,t) &= \intz f_J(\w,t)\, d\wpos
& m_J^\neg(t)&= \frac1{\rho_J(t)}\ \intzz \wpos\, f_J(\w,t)\, \dw
\end{align*}
denote the marginals densities, i.e., the fractions of the population that 
belongs to compartment^^>$J\in \C$ 
with positive and negative opinion, respectively, at time $t\ge 0$, while in 
the second column we denote the mean relative to the positive and to the 
negative opinion, respectively. Finally, 
\[
\rho_J(t) = \int_{[0,1]^2} 
f_J(\w,t)\, \dw
\]
is the total mass fraction of agents in the compartment^^>$J$.

\begin{figure}\centering
\includegraphics[width=0.6\linewidth]{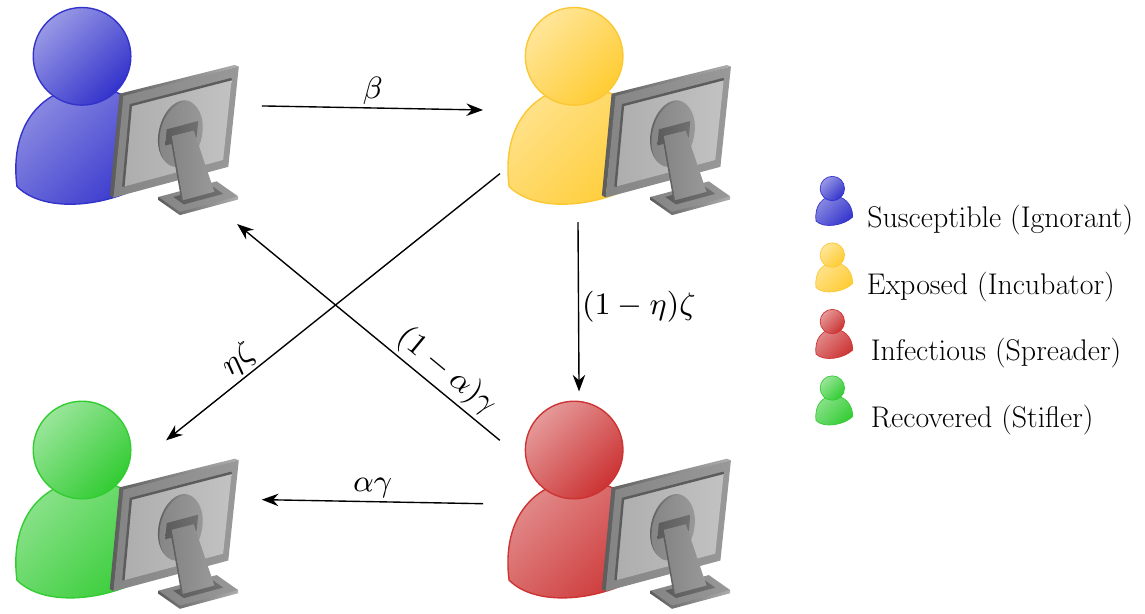}
\caption{Compartmental dynamic and parameters in the fake news SEIR 
model^^>\eqref{eq:SEIRode}: people get exposed to fake news with a contact rate $\beta$ with infected individuals, after a latency 
period $1/\zeta$ with probability $1-\eta$ they start spreading it until they finally stop after an average time $1/\gamma$ 
and become \lq immunized\rq\ or uninterested in it, thus removing themselves from the 
dissemination dynamics with probability $\alpha$.}
\label{fig:seir-scheme}
\end{figure}

\begin{figure}
\centering
\includegraphics[width=\linewidth]{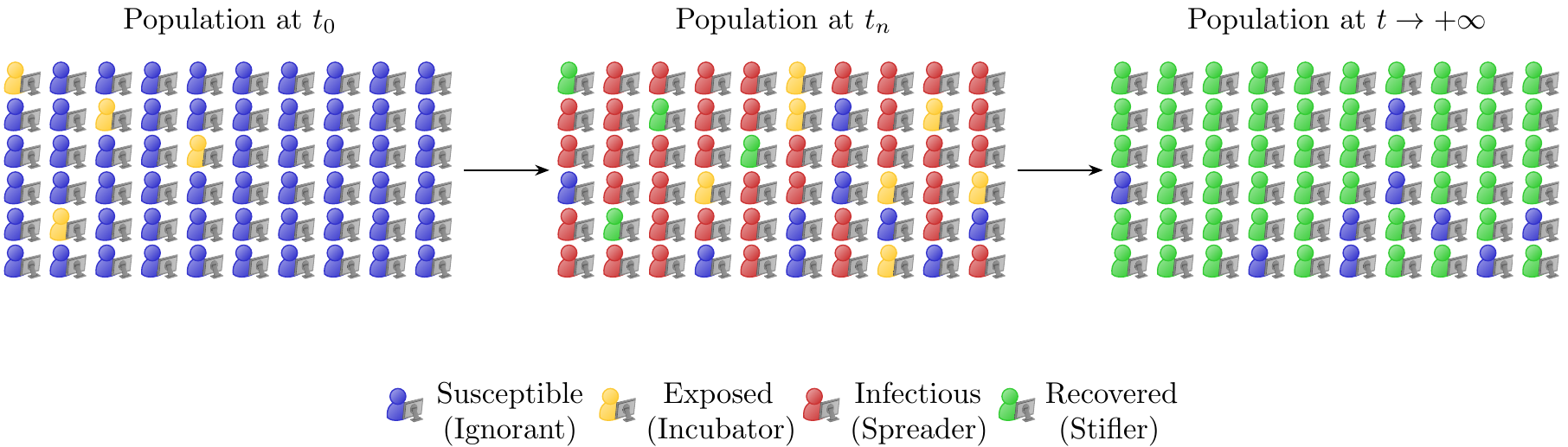}
\caption{Evolution in time of the dissemination of fake news within a 
population for model^^>\eqref{eq:SEIRode}. For large times the fake-news infection disappears and the population is composed only by susceptible and recovered individuals.
}
\label{fig:seir-evo}
\end{figure}

When the fake-news dynamic is independent from the opinion of individuals it 
follows the simple system of ordinary differential equations^^>\cite{fp22}
\(
\begin{gathered}
\left\lbrace
\begin{aligned}
\der{\rho_S(t)}{t} &= -\beta \rho_S(t)\rho_I(t)+(1-\alpha)\gamma\rho_I(t)\\
\der{\rho_E(t)}{t} &=  \beta \rho_S(t)\rho_I(t) -\zeta \rho_E(t)\\
\der{\rho_I(t)}{t} &= (1-\eta)\zeta \rho_E(t) - \gamma \rho_I(t)\\
\der{\rho_R(t)}{t} &= \eta\zeta \rho_E(t)+\alpha\gamma \rho_I(t)
\end{aligned}
\right.
\end{gathered}
\label{eq:SEIRode}
\)
with $\rho_S(t) + \rho_E(t) + \rho_I(t) + \rho_R(t) = 1$.
We will refer to it as a SEIR model for fake-news spreading. Basically, susceptible agents get exposed 
at a rate that is proportional to the probability of them interacting with an 
active spreader (i.e., an infectious individual). Once they are exposed, they 
wait for an average time $1/\zeta$ and start disseminate the fake news with probability $1-\eta$. 
After an average time $1/\gamma$, they stop doing so and are removed 
permanently from the 
dynamics with probability $\alpha$. 
A schematic depiction of the dynamics is showed in Fig^^>\ref{fig:seir-scheme}, whereas  
in Fig^^>\ref{fig:seir-evo} an example of evolution within a closed population 
is sketched. In the sequel, for simplicity, we assume $\eta=0$, $\alpha=1$ so that the exposed individuals, after the
latency period, always start to spread the fake-news and after an average time spreaders are
permanently removed from the dynamic. 

If we combine the dissemination dynamics with the opinion formation process 
described in the previous section we obtain the following mean-field model

\begin{align}
\nonumber
\pd{f_S(\w,t)}{t} &= -K(f_S, f_I)(\w,t) + \lampos_S\pd{}{\wpos}[\bigl(\wpos -  m_\pos (t)\bigr) f_S(\w,t)] \\
\label{eq:SEIRFP1}
&\hphantom{{}=} + \lamneg_S\pd{}{\wneg}[\bigl(\wneg -  m_\neg  (t)\bigr) f_S(\w,t)] \\
\nonumber
&\hphantom{{}=} + \frac{\sigma_{\pos,S}^2}{2} 
\pd{^2}{\wpos^2}\bigl(D(\wpos)^2 f_S(\w,t)\bigr) + 
\frac{\sigma_{\neg,S}^2}{2}
\pd{^2}{\wneg^2}\bigl(D(\wneg)^2 f_S(\w,t)\bigr),\\
\nonumber
\pd{f_E(\w,t)}{t} &=  K(f_S, f_I)(\w,t) - \zeta(\w) f_E(\w,t)
                     + \lampos_E\pd{}{\wpos}[\bigl(\wpos - 
  m_\pos (t)\bigr) f_E(\w,t)]\\
 \label{eq:SEIRFP2}
&\hphantom{{}=} + \lamneg_E\pd{}{\wneg}[\bigl(\wneg - 
  m_\neg (t)\bigr) f_E(\w,t)] + \frac{\sigma_{\pos,E}^2}{2} 
\pd{^2}{\wpos^2}\bigl(D(\wpos)^2 f_E(\w,t)\bigr)\\
\nonumber
&\hphantom{{}=}+ \frac{\sigma_{\neg,E}^2}{2}
\pd{^2}{\wneg^2}\bigl(D(\wneg)^2 f_E(\w,t)\bigr),\\
\nonumber
\pd{f_I(\w,t)}{t} &= \zeta(\w) f_E(\w,t) - \gamma(\w)f_I(\w,t)
                     + \lampos_I\pd{}{\wpos}[\bigl(\wpos - 
  m_\pos (t)\bigr) f_I(\w,t)]\\
 \label{eq:SEIRFP3}
&\hphantom{{}=}+ \lamneg_I\pd{}{\wneg}[\bigl(\wneg - 
  m_\neg (t)\bigr) f_I(\w,t)] + \frac{\sigma_{\pos,I}^2}{2} 
\pd{^2}{\wpos^2}\bigl(D(\wpos)^2 f_I(\w,t)\bigr)\\
\nonumber
&\hphantom{{}=} + \frac{\sigma_{\neg,I}^2}{2}
\pd{^2}{\wneg^2}\bigl(D(\wneg)^2 f_I(\w,t)\bigr),\\
\nonumber
\pd{f_R(\w,t)}{t} &= \gamma(\w) f_I(\w,t)
                     + \lampos_R\pd{}{\wpos}[\bigl(\wpos - 
  m_\pos (t)\bigr) f_R(\w,t)] \\
 \label{eq:SEIRFP4}
&\hphantom{{}=} + \lamneg_R\pd{}{\wneg}[\bigl(\wneg - 
  m_\neg (t)\bigr) f_R(\w,t)] \\
\nonumber
&\hphantom{{}=} + \frac{\sigma_{\pos,R}^2}{2} 
\pd{^2}{\wpos^2}\bigl(D(\wpos)^2 f_R(\w,t)\bigr) + 
\frac{\sigma_{\neg,R}^2}{2}
\pd{^2}{\wneg^2}\bigl(D(\wneg)^2 f_R(\w,t)\bigr),
\end{align}
where 
\(
 m_\pos = \sum_{J \in \C} \rho_J m_J^\pos,\qquad 
 m_\neg = \sum_{J \in \C} \rho_J m_J^\neg.
\) 
The functional
\(
K(f_S,f_I)(\w,t)  =f_S(\w,t) \intzz \kappa(\w*) f_I(\w*,t)\, \dw*
\)
is the local incidence rate of interactions between susceptible and infectious individuals, where $\kappa(\w)$ is a contact function which 
measures the impact of the opinion in the dissemination of fake-news. A 
simplifying assumption is that $\kappa(\phv,\phv)$ is separable in the two 
variables, i.e., $\kappa(\w) = \beta k(\wpos)\bar k (\wneg)$, with $\beta > 0$ 
a constant. A choice of particular interest would be one in which 
$\kappa(\phv)$ is a function of the sole variable^^>$\wpos$ (respectively, 
$\wneg$).
System^^>\eqref{eq:SEIRFP1}-\eqref{eq:SEIRFP4} needs to be complemented by the no flux boundary conditions for all $J \in \C$
\(\label{eq:boundcondcompartment}
\begin{aligned}
\lampos_J\bigl(\wpos -  m_\pos(t)\bigr) f_J(\w,t) +
\frac{\sigma^2_{+,J}}{2}\pd{}{\wpos} \bigl(D(\wpos)^2 f_J(\w,t)\bigr)  = 0,\quad {{\rm on\,\,} w_+=0,1},\\
\lamneg_J\bigl(\wneg -  m_\neg(t)\bigr) f_J(\w,t) +,
\frac{\sigma^2_{-,J}}{2}\pd{}{\wneg} \bigl(D(\wneg)^2 f_J(\w,t)\bigr)  = 0,\quad {{\rm on\,\,} w_-=0,1}.
\end{aligned}
\)
Note that, if the alignment rates $\lambda^\pm_J = \lambda^\pm$ independent 
from $J\in \mathcal{C}$, as a consequence of the above boundary conditions and 
the choice of the diffusion function \eqref{eq:diff}, the quantities ${m}_+(t)$ 
and ${m}_-(t)$ are conserved in time. 
 
\subsection{Stationary marginal densities}

If in system^^>\eqref{eq:SEIRFP1}-\eqref{eq:SEIRFP4}  we choose a constant 
function^^>$\kappa(\phv,\phv) \equiv \beta > 0$ as well as constant 
epidemiological parameters we obtain again system^^>\eqref{eq:SEIRode} with $\alpha=\eta=1$ by 
integrating in the variable^^>$\w$. As a consequence, classical results in epidemiology^^>\cite{heth00}
guarantee that when $t \to \infty$ the fake-news spreading vanishes and we 
have both $\rho_E(t) \to 0$ and $\rho_I(t) \to 0$. Moreover, $\rho_S(t)\to \rho_S^\infty$ and
$\rho_R(t)\to \rho_R^\infty = 1 -\rho_S^\infty$ where $\rho_S^\infty$ solves
\[
\log\left(\frac{\rho_S}{\rho_S(0)}\right) = \frac{\beta}{\gamma} (1-\rho_S^\infty).
\] 
Let us denote with $m_+^\infty$, $m_-^\infty$ the large time behavior 
Similarly the evolutions of the first moments are obtained by integrating in $\w$ after multiplication for $w_\pm$ to obtain 
\(
\begin{gathered}
\left\lbrace
\begin{aligned}
\frac{d}{dt}(\rho_S(t)m^\pm_S(t)) &= -\beta \rho_S(t)m^\pm_S(t)\rho_I(t)-\lambda^\pm_S\rho_S( m^\pm_S-{m}_\pm)\\
\frac{d}{dt}(\rho_E(t)m^\pm_E(t)) &=  \beta \rho_S(t)m^\pm_S(t)\rho_I(t) -\zeta \rho_E(t)-\lambda^\pm_E\rho_E( m^\pm_E-{m}_\pm)\\
\frac{d}{dt}(\rho_I(t)m^\pm_I(t)) &= \zeta \rho_E(t)m^\pm_E(t) - \gamma \rho_I(t)m^\pm_I(t)-\lambda^\pm_I\rho_I( m^\pm_I-{m}_\pm)\\
\frac{d}{dt}(\rho_R(t)m^\pm_R(t)) &= \gamma \rho_I(t)m^\pm_I(t)-\lambda^\pm_R\rho_R( m^\pm_R-{m}_\pm).
\end{aligned}
\right.
\end{gathered}
\label{eq:SEIRodemean}
\)
So that for large times we have $m_\pm(t)\to m_\pm^\infty$ and $m^\pm_S=m^\pm_R=m_\pm^\infty$.

Then, if we integrate \eqref{eq:SEIRFP1}-\eqref{eq:SEIRFP4}  with respect to $\wneg$ in the same way 
we did for equation^^>\eqref{eq:FPopinion}, using the boundary conditions \eqref{eq:boundcondcompartment} we have that at the stationary state the marginals for the 
positive opinion satisfy
\[
\begin{aligned}
\pd{}{\wpos} \left[
\bigl(\lampos_S\wpos - {m}^\infty_\pos\bigr) 
\gpj[S]^\infty(\wpos)\right] + 
\frac{\sigma_{\pos,S}^2}{2} \pd{^2}{\wpos^2} \left[
\wpos(1-\wpos) \gpj[S]^\infty(\wpos)\right] &= 0,\\
\pd{}{\wpos} \left[
\bigl(\lampos_R\wpos - {m}^\infty_\pos\bigr) 
\gpj[R]^\infty(\wpos)\right] + 
\frac{\sigma_{\pos,R}^2}{2} \pd{^2}{\wpos^2} \left[
\wpos(1-\wpos) \gpj[R]^\infty(\wpos)\right] &= 0,
\end{aligned}
\]
which provide the stationary distributions
\(\label{eq:steadycompartment}
\begin{aligned}
\gpj[S]^\infty(\wpos) &=  
\rho_S^\infty C^\pos_S 
\wpos^{{m}^\infty_\pos/\mu^+_{S}-1}
(1-\wpos)^{(1-{m}^\infty_\pos)/\mu^+_{S}-1},\\
\gpj[R]^\infty(\wpos) &=  
(1 - \rho_S^\infty) C^\pos_R 
\wpos^{{m}^\infty_\pos/\mu^+_{R}-1}
(1-\wpos)^{(1-{m}^\infty_\pos)/\mu^+_{R}-1}
\end{aligned}
\)
where $\mu^+_{S} = \lambda^+_{S}/\sigma^2_{\pos,S}$, $\mu^+_{R} = \lambda^+_{R}/\sigma^2_{\pos,R}$ and $C^+_{S}$, $C^+_{R}$ are normalization constants.
 
 Introducing analogous hypotheses, 
we obtain the same result for the total marginal density of negative opinions
\(
\label{eq:steadycompartmentneg}
\begin{aligned}
h_S^\infty(\wneg) &=  
\rho_S^\infty C^\neg_S 
\wneg^{{m}^\infty_\neg/\mu^-_{S}-1}
(1-\wneg)^{(1-{m}^\infty_\neg)/\mu^-_{S}-1},\\
h_R^\infty(\wneg) &=  
(1 - \rho_S^\infty) C^\neg_R 
\wneg^{{m}^\infty_\neg/\mu^-_{R}-1}
(1-\wneg)^{(1-{m}^\infty_\neg)/\mu^-_{R}-1}
\end{aligned}
\)
with $\mu^-_{S} = \lambda^-_{S}/\sigma^2_{-,S}$, $\mu^-_{R} = \lambda^-_{R}/\sigma^2_{-,R}$ and $C^-_{S}$, $C^-_{R}$ normalization constants.

\begin{figure}
\newsavebox\figdue
\savebox\figdue{\includegraphics[width=0.475\linewidth]{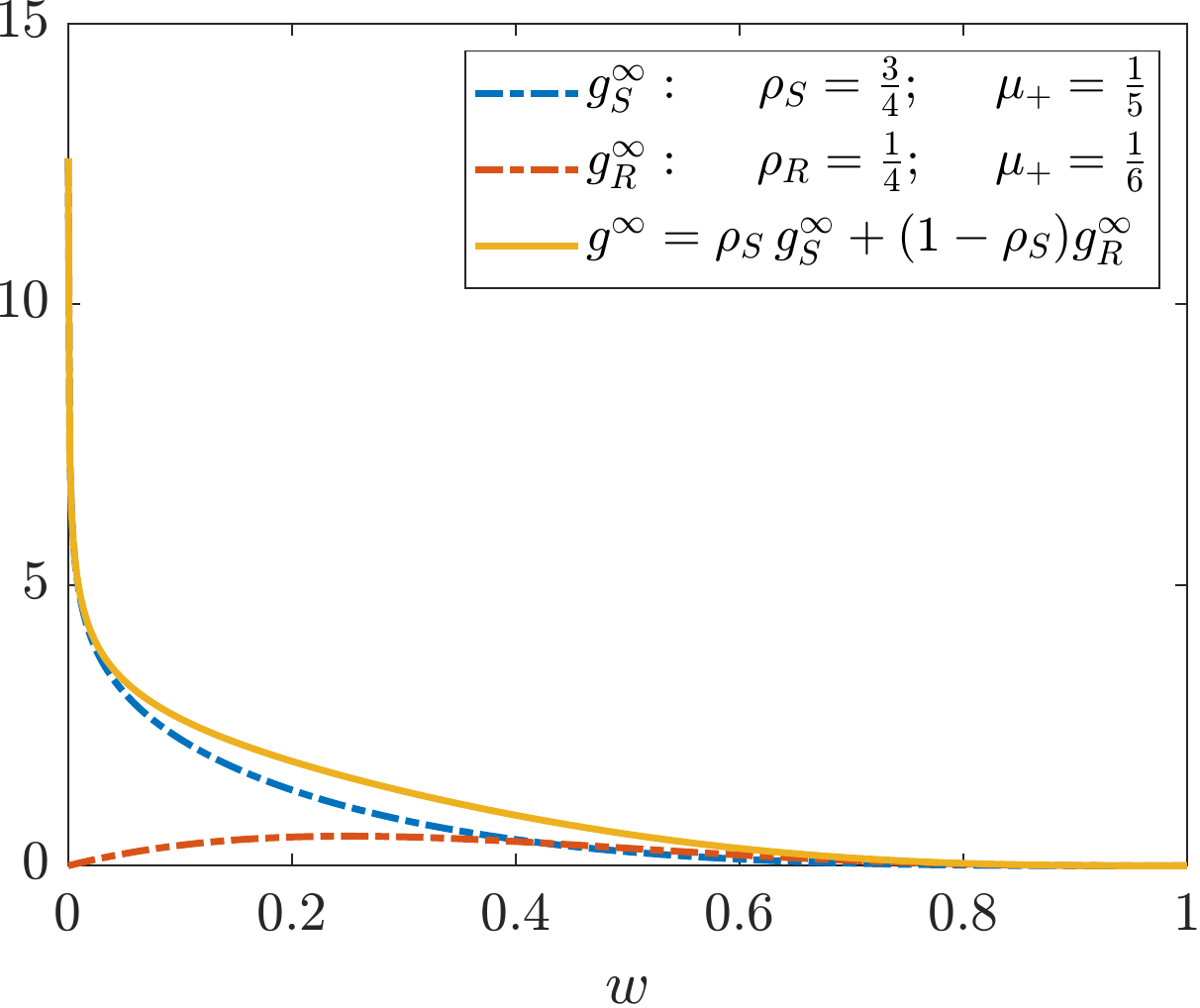}}
\usebox\figdue\hfill
\includegraphics[width=0.475\linewidth,height=\ht\figdue]{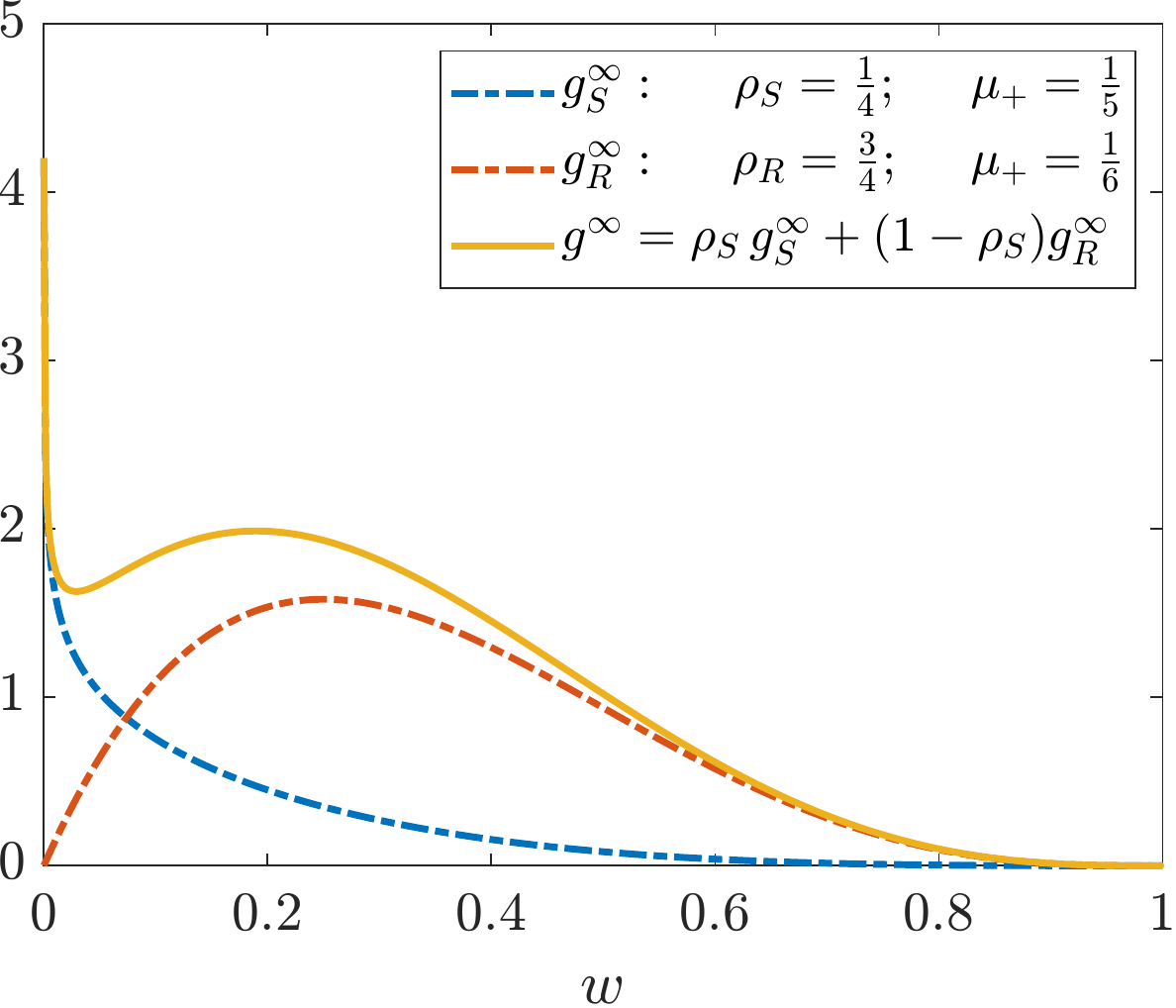}
\caption{Examples of stationary solutions obtained from \eqref{eq:steadycompartment} with different choices of parameters $\rho_S$, $\rho_R$, ${m}_+$, $\lambda^+_S$, $\lambda^+_R$, $\sigma_{+,S}$ and $\sigma_{-,S}$. We can see that the resulting function can have both a unimodal character (left) or a bimodal behavior (right).}
\label{fig:betamixture}
\end{figure}

This means that, depending on the given regime of parameters, since the total 
marginal density for the positive or negative opinions is the mixture of two Beta 
distributions, it can be unimodal or bimodal (see Fig \ref{fig:betamixture}).

\section{Application to Covid-19 vaccination hesitancy}\label{sec:data}
As we mentioned in the introduction, the proposed model is suitable to describe 
the evolution of opinions within closed online communities where data are collected through
a suitable use of NLP techniques like sentiment analysis. Here, we present the 
particular case of selected groups of people which shared preoccupation for the 
Italian vaccination campaign in response to the SARS-CoV-2 
pandemic^^>\cite{rubin22}. More in 
details, they all share, to a certain extent, \emph{vaccination hesitancy}.
 
\subsection{Data collection}

The groups we focus on are composed of users of online group chats on Telegram. 
We collected a total of 4077 posts from six different 
chats, from August 20\th, 2021 to February 27\th, 2022. They all focus on 
vaccination, from various perspectives and scopes, which refer to different 
social groups the users belong to.

References to conspiracy theories, plain misinformation, rage bursts and 
mockery, all mix with sincere pleas of explanations on vaccination as well as 
other measures employed by the former Italian government to combat and contain 
the effect of the SARS-CoV-2 pandemic on both the national healthcare and 
economy.

Using techniques proper of the framework of \emph{sentiment 
analysis}^^>\cite{yue19,nemes21,melton22,piedrahita21,nicola18}, we 
assigned to each post a pair $(\bar w_\pos, \bar w_\neg) \in [0,1]^2$ of 
scores, respectively positive and negative which reflect how good or bad the 
opinion of the user might be in that instant of time.

Scores are assigned in an automatic fashion by a pre-trained model which 
analyzes the content of each post. The following is a post from our dataset, 
with the English translation aside.
\begin{quote}
\begin{minipage}{0.45\linewidth}
Ma tanto anche se a queste persone diciamo non fatevi più altre dosi che vi 
fanno male, non ci ascoltano. Ormai per loro siamo noi i cattivi, e non quelli 
che veramente sono i cattivi. Ormai non hanno in mente altro. Pensano solo alla 
prossima dose e ai cattivi no vax.
\end{minipage}\hfil
\begin{minipage}{0.45\linewidth}
Still even if we tell this people don't take any more doses, that they hurt 
you, they won't listen. At this point to them it's us the bad guys and not the 
ones that are the bad guys for real. By now they have nothing else in mind. 
They just think about the next dose and to the bad no-vaxs.
\end{minipage}
\end{quote}
This post was evaluated as a score pair of $(0.055934787,0.820981)$. 
This one instead scored a pair $(0,31506833, 0.69410014)$.
\begin{quote}
\begin{minipage}{0.45\linewidth}
Io di Paragone non mi fido\ldots cmq io la mia battaglia la faccio qui. E siamo 
davvero na marea. Non so come e in Italia ma qui pian piano la gente si sta 
svegliando, anche i vaccinati si stanno unendo a noi.
\end{minipage}\hfil
\begin{minipage}{0.45\linewidth}
I don't trust Paragone [former Italian politician, authors' note]\ldots however 
I fight 
my battle here. And we really are a ton. I don't know how and in Italy but here 
slowly but surely people are waking up, even the vaccinated are joining us. 
\end{minipage}
\end{quote}
As a last example, we report one of the few posts that were originally written 
in English (score of $(0.028655171,0.045119375)$).
\begin{quote}
\textbf{\textsc{truth revealed}}: FAUCI just confess on a live stream with Mark 
Zuckerberg that vaccination actually may cause the problem.
\end{quote}
This post is revealing of some of the issues involved in using software-based 
sentiment analysis techniques: the form may be neutral and plain, but its 
content arguably is. Also the small caps text is typical of the sensationalist 
tones affine with conspiracy theories and fake news in general.

In spite of sentiment analysis having become spread both in academic 
and corporate works^^>\cite{piedrahita21,klimiuk21,agustiningsih21}, its 
evaluation is not free from risk: since NLP is a relatively young discipline which faces lots of 
challenging tasks, there are no current one-solves-all approaches 
for parsing human-produced syntax in a robust way. Besides, online chats might 
not be the most suited environment for unambiguous, error-free communication, 
not to mention the use of non-verbal means, such as non-plain-text characters 
(emojis, for instance) to express emotions and concepts which necessarily would 
go undetected by a not instructed software. Hence, the evaluation of the 
records in our datasets comes with inherent 
uncertainties. Here we do not try to quantify these uncertainties, we 
refer to^^>\cite{apz22} for related approaches to uncertain data
in compartmental models.

In the following, when interfacing data with our model, we always considered 
aggregate data, i.e., scores gathered for posts from every group chat combined 
into a unique dataset. Moreover, if not otherwise specified, we always 
discretized the dataset into a grid of $20\times20$ bins. 
\begin{figure}[tb]
\centering
\includegraphics[width=0.5\linewidth]{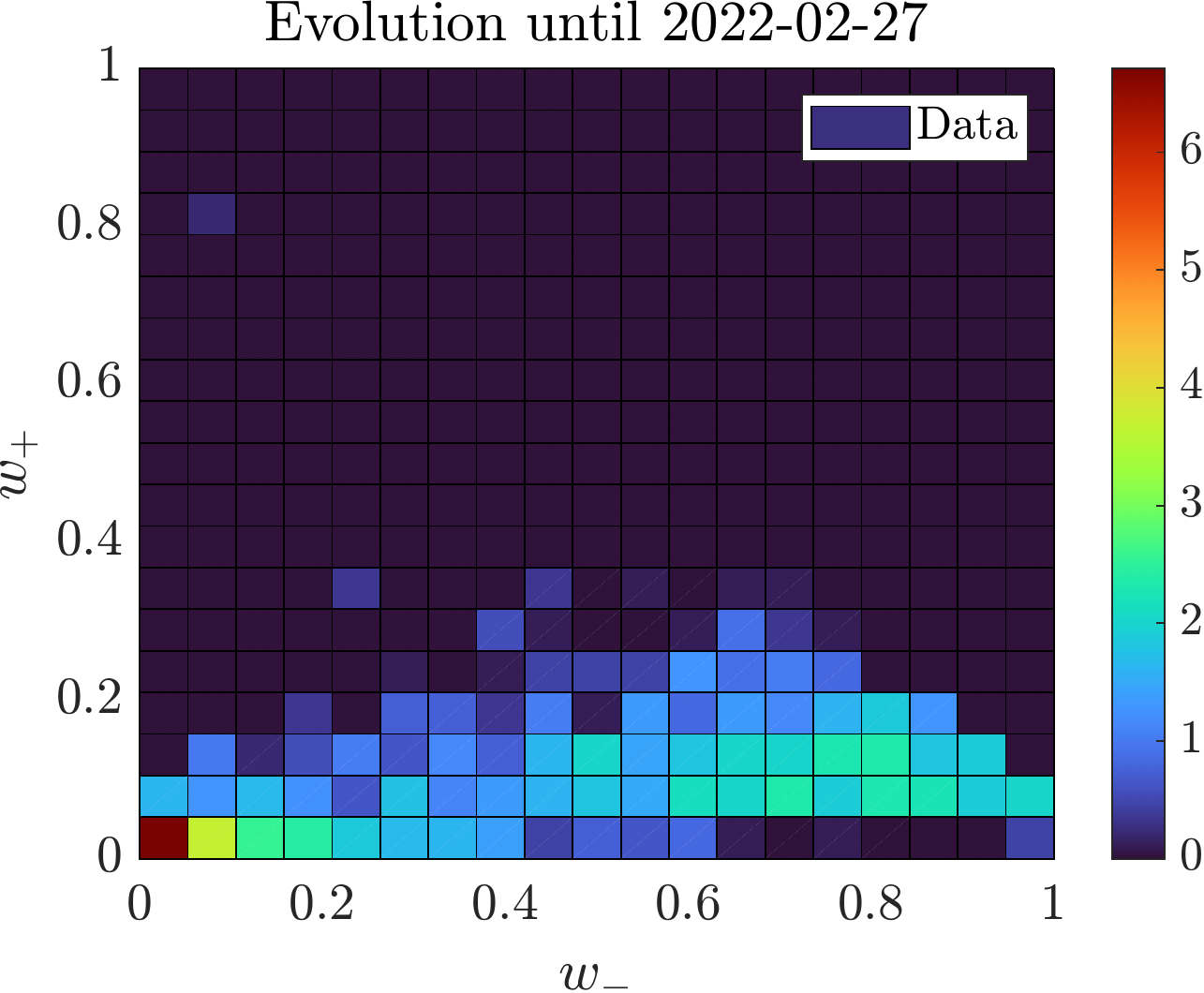}%
\includegraphics[width=0.5\linewidth]{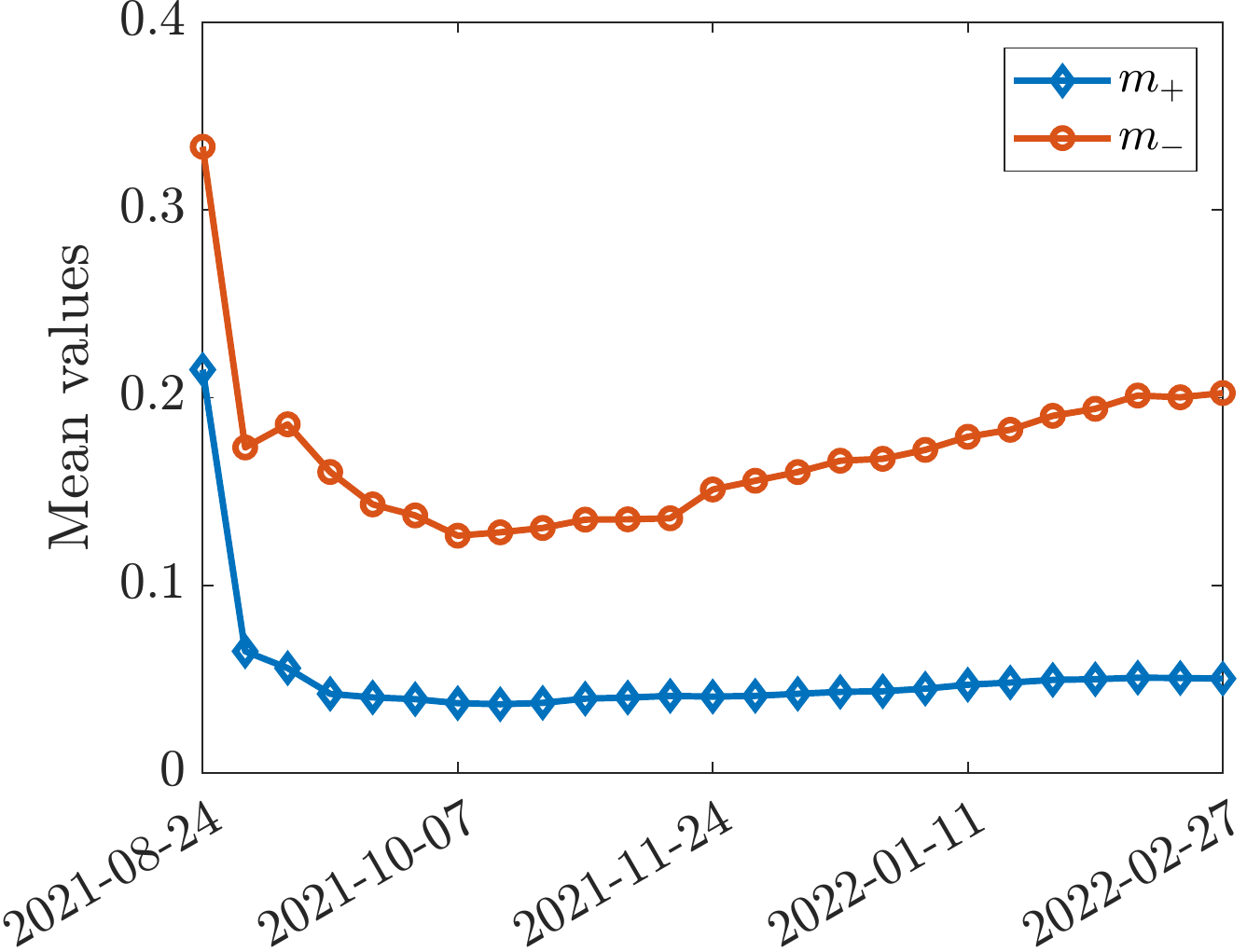}
\caption{Left: final time snapshot of the dataset. The base 2-logarithm of the 
data is used for coloring in order to better appreciate the differences in 
magnitude. Two main concentration regions are clearly visible: the stronger 
one, around the origin, with the highest peak and the lowest local variance, 
and the weaker one, in the bottom right region of the surface plot. Right: time 
evolution of mean positive and negative opinion. While the mean positive 
opinion quickly converges toward an equilibrium point, the negative one 
presents an increasing trend after a brief decreasing phase.}
\label{fig:data}
\end{figure}
The main peculiarity of the dataset is depicted in Fig.^^>\ref{fig:data}: 
at the end of the evolution period, a significant concentration of people with 
strong negative opinion and essentially neutral positive opinion has formed. 
This is precisely the kind of clustering polarization that we mentioned in 
previous sections: here the dataset is showing a clear instance of bimodal 
distribution. Moreover, as the evolution of the positive and negative mean 
opinions shows, this bimodal distribution is the outcome of a polarizing trend 
across the population, which, interestingly, involves only the negative 
opinions.

\begin{figure}
\centering
\includegraphics[width=0.5\linewidth]{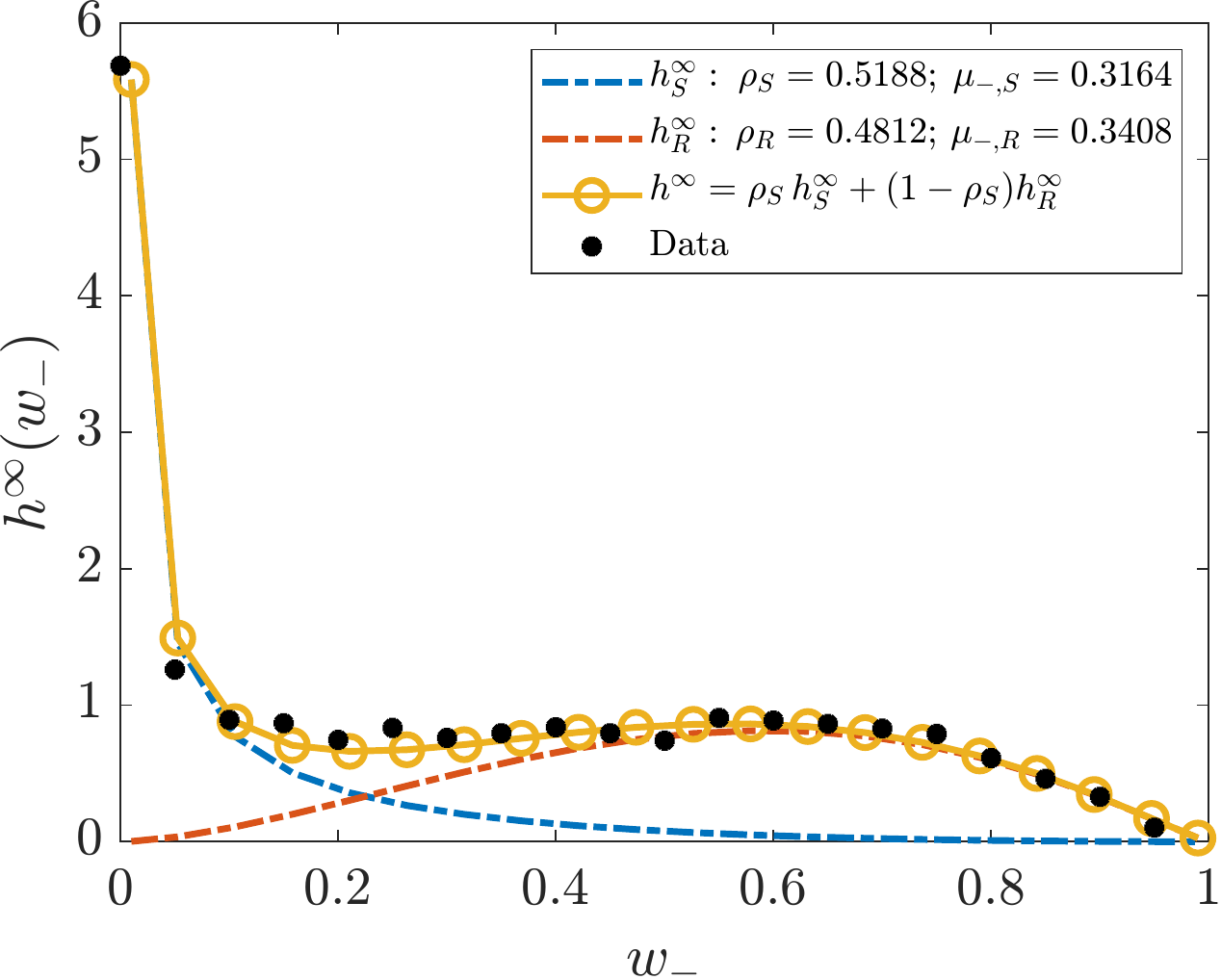}
\caption{Fitting of equilibrium data: marginal of the negative opinion with a 
convex combination of the marginal distribution for susceptible and recovered 
individuals, respectively. The fitting parameters are shown and for 
ease of retrieval are also reported in Table^^>\ref{tab:fitpar}.}
\label{fig:betafit}
\end{figure}

Let us focus on the marginal of the negative opinion at the final time 
snapshot. If we make the \emph{ansatz} that it can be well approximated by a 
suitable convex combination of two Beta distributions, we obtain the plot 
showed in Fig.^^>\ref{fig:betafit}. We report in Table^^>\ref{tab:fitpar} the 
fitting parameters, obtained as the solution of the problem
\(\label{eq:fitprob}
\min_{\substack{\overline \mu_{-,S},\ \mu_{-,R},\\ m_-^\infty, \ 
\rho_S^\infty}}
\|\rho_S^\infty h_S^\infty(\wneg;  m_-^\infty, \mu_{-,S}) + 
 (1-\rho_S^\infty)h_R^\infty(\wneg;  m_-^\infty, \mu_{-,R}) -  
\underline h^\infty(\wneg)\|_2,
\)
where $\underline h^\infty(\phv)$ is the marginal distribution of the 
negative opinion extrapolated from the last recorded time snapshots.

\begin{table}[b]
\centering
\begin{tabular}{*8{c}}
\toprule
\multicolumn{8}{c}{Parameters}\\
\midrule
$\rho_S$ & $m_\infty^- $ & $\mu_{-,S}$  & $\mu_{-,R}$ &
$\lambda^+_S$ & $\lambda^+_R$ & $\lambda^-_S$ & $\lambda^-_R$\\
\midrule
0.5188 & 0.0793 & 0.3164 & 0.3408 &
6.0000 & 1.500  & 4.0000 & 0.4700\\
\bottomrule
\end{tabular}
\caption{Fitting parameters obtained as solution of minimization 
problem^^>\eqref{eq:fitprob}.}
\label{tab:fitpar}
\end{table}
\subsection{Simulation results} 
The next section is devoted to compare the evolution provided by the 
model^^>\eqref{eq:SEIRFP1}--\eqref{eq:SEIRFP4} and the one provided by the 
data. The model has been 
calibrated with the parameters in Table^^>\ref{tab:fitpar}. For what concerns 
the epidemiological coefficients, they were chosen in order to achieve total 
masses at equilibrium that were compatible with the quantity^^>$\rho_S$ in 
Table^^>\ref{tab:fitpar}, which gives the mass fractions of the two Beta 
distributions that concur to provide the steady state for the marginal negative 
density. 

Computing the numerical evolution of the model requires a careful discretization of the system in order to keep
high accuracy when describing the stationary solutions. To this aim we adopt the steady state preserving approach 
devised for Fokker-Planck equations in^^>\cite{pareschi18} by extending it to systems in our multidimensional case.

To this aim, system^^>\eqref{eq:SEIRFP1}-\eqref{eq:SEIRFP4} was split both in time and opinion space. To describe the 
splitting, let us rewrite the system as follows
\[
\partial_t \mathbf f(\w,t) = \mathbf F_+[\mathbf f(\w,t)](\wpos,t) + \mathbf 
F_-[\mathbf f(\w,t)](\wneg,t) + \mathbf P[\mathbf f(\w,t)](\w,t),
\]
where the bold operators are vector valued such as
\[
\begin{aligned}
\mathbf F_+[\mathbf f(\w,t)](\wpos,t) &= 
\left[\pd{}{\wpos}[\bigl(\lampos_J\wpos - 
 m_\pos (t)\bigr) f_J(\w,t)] + \frac{\sigma_{\pos,J}^2}{2} 
\pd{^2}{\wpos^2}\bigl(D(\wpos)^2 f_J(\w,t)\bigr)\right]_J, \quad J \in \C\\
\mathbf F_-[\mathbf f(\w,t)](\wneg,t) &=
\left[\pd{}{\wneg}[\bigl(\lamneg_J\wneg - 
 m_\neg (t)\bigr) f_J(\w,t)] + \frac{\sigma_{\neg,J}^2}{2} 
\pd{^2}{\wneg^2}\bigl(D(\wneg)^2 f_J(\w,t)\bigr)\right]_J, \quad J \in C\\
\mathbf P[\mathbf f(\w,t)](\w,t) &= \bigl[-K(f_S, f_I); K(f_S, f_I)  - \zeta 
f_E; \zeta f_E - \gamma f_I; \gamma f_I  \bigr](\w,t).
\end{aligned}
\]
Then, if we discretize the time domain with a time step of size $\Delta t>0$ 
and we denote by $\mathbf f^n(\w)$ an approximation of $\mathbf f(\w,n\Delta 
t)$, the (first-order) time splitting method consists in solving in the time interval $[0,\Delta t]$ the following sequence of problems
\begin{gather}
\text{Evolve positive opinions $\Rightarrow$}
\left\lbrace
\begin{aligned}
\pd{\mathbf f^\dag}{t} &= \mathbf F_+[\mathbf f^\dag],\\
\mathbf f^\dag(\w,0) &= \mathbf f^n(\w),\qquad 
\end{aligned}
\right.\\
\text{Evolve negative opinions $\Rightarrow$}
\left\lbrace
\begin{aligned}
\pd{\mathbf f^{\dag\dag}}{t} &= \mathbf F_-[\mathbf f^{\dag\dag}],\\
\mathbf f^{\dag\dag}(\w,0) &= \mathbf f^\dag(\w,\Delta t),
\end{aligned}
\right.\\
\text{Evolve fake-news spreading $\Rightarrow$}
\left\lbrace
\begin{aligned}
\pd{\mathbf f^{\dag\dag\dag}}{t} &= \mathbf P[\mathbf f^{\dag\dag\dag}],\\
\mathbf f^{\dag\dag\dag}(\w,0) &= \mathbf f^{\dag\dag}(\w,2\Delta t),
\end{aligned}
\right.
\end{gather}
and finally set $\mathbf f^{n+1}(\w) = \mathbf f^{\dag\dag\dag}(\w,\Delta t)$. 
Higher 
order splitting can 
be constructed as well (see^^>\cite{pareschi18} and the references therein). 

\begin{figure}[t]
\centering
\includegraphics[width=0.4\linewidth]{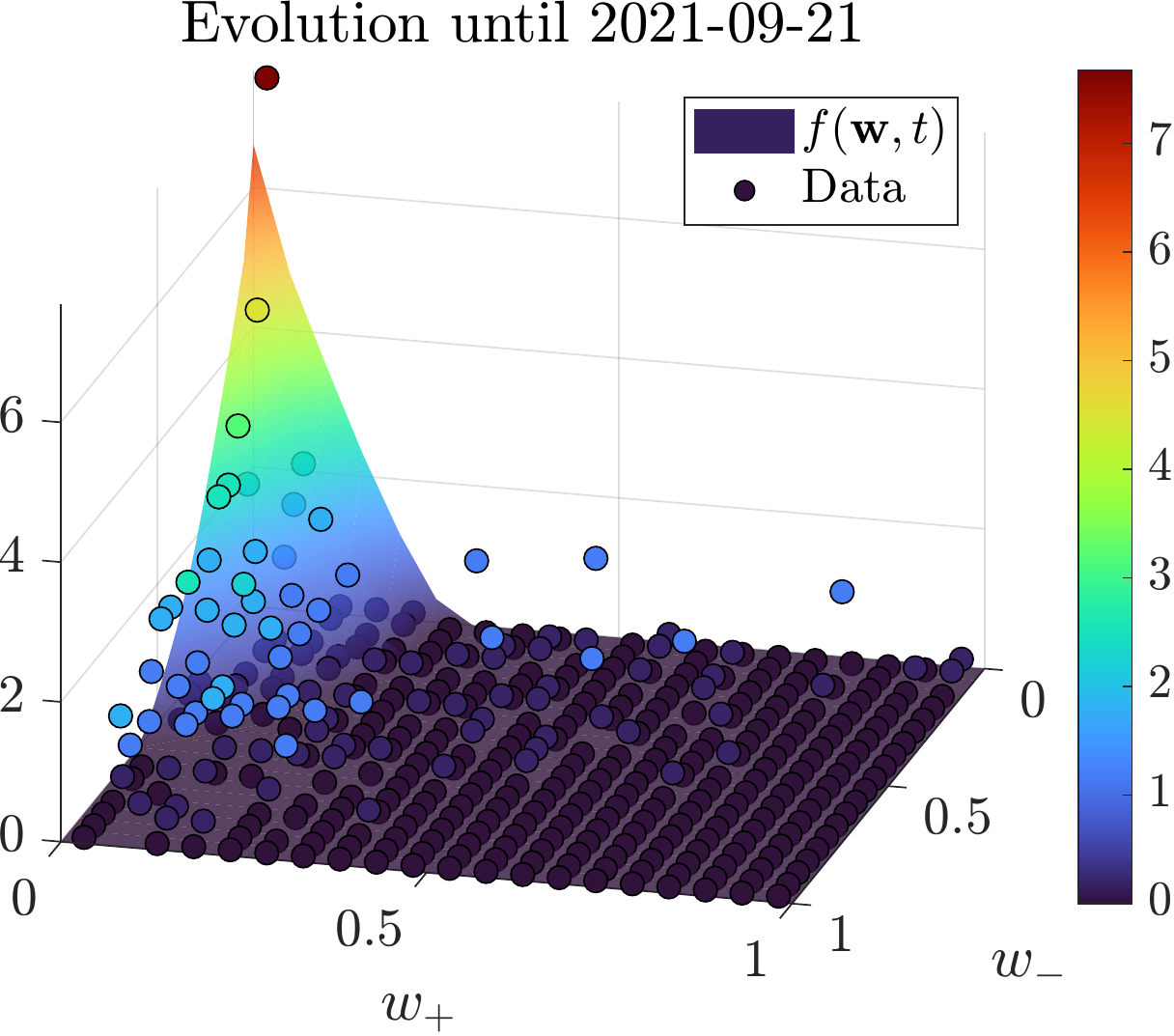}\hskip 0.5cm%
\includegraphics[width=0.4\linewidth]{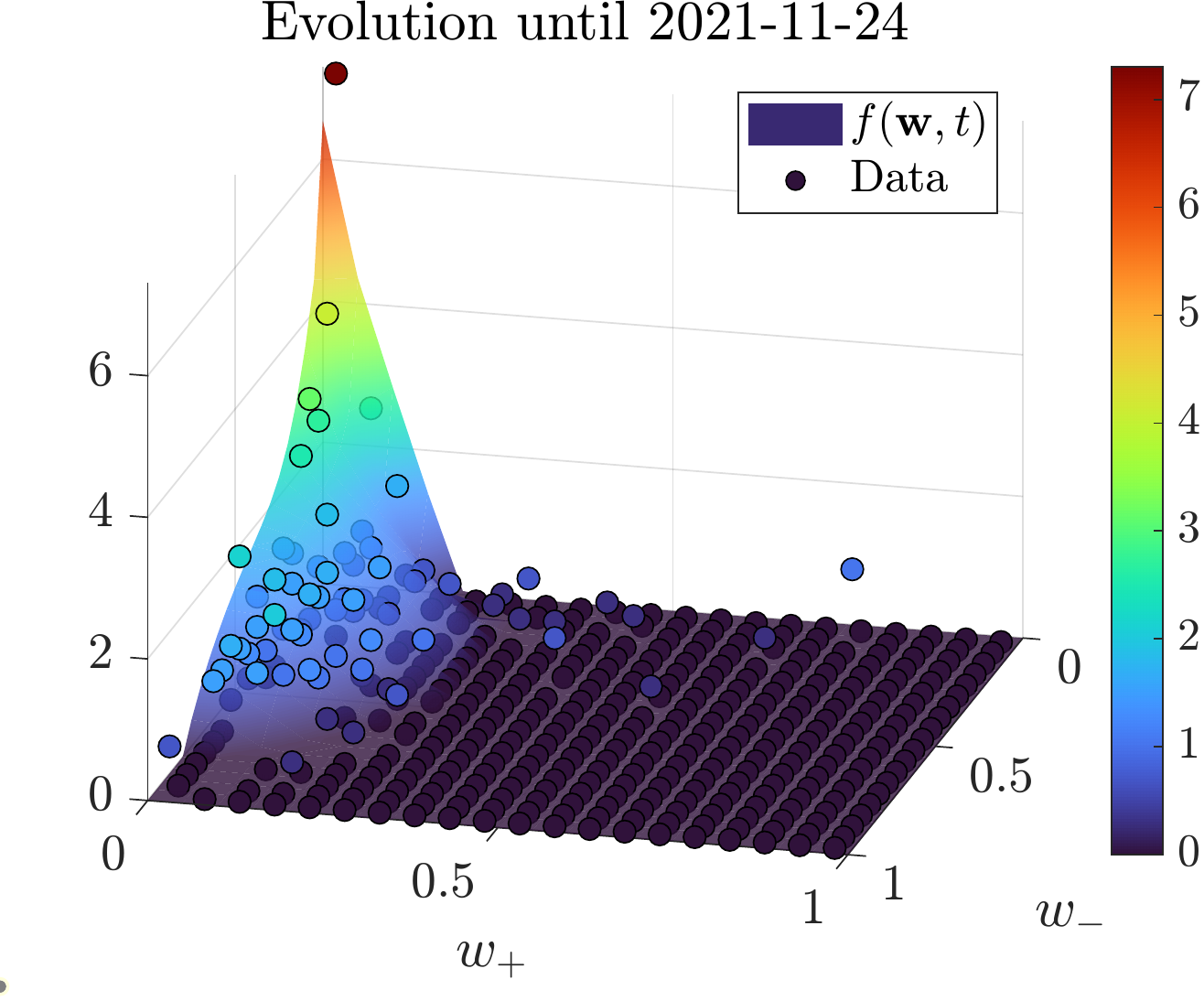}\\%
\includegraphics[width=0.4\linewidth]{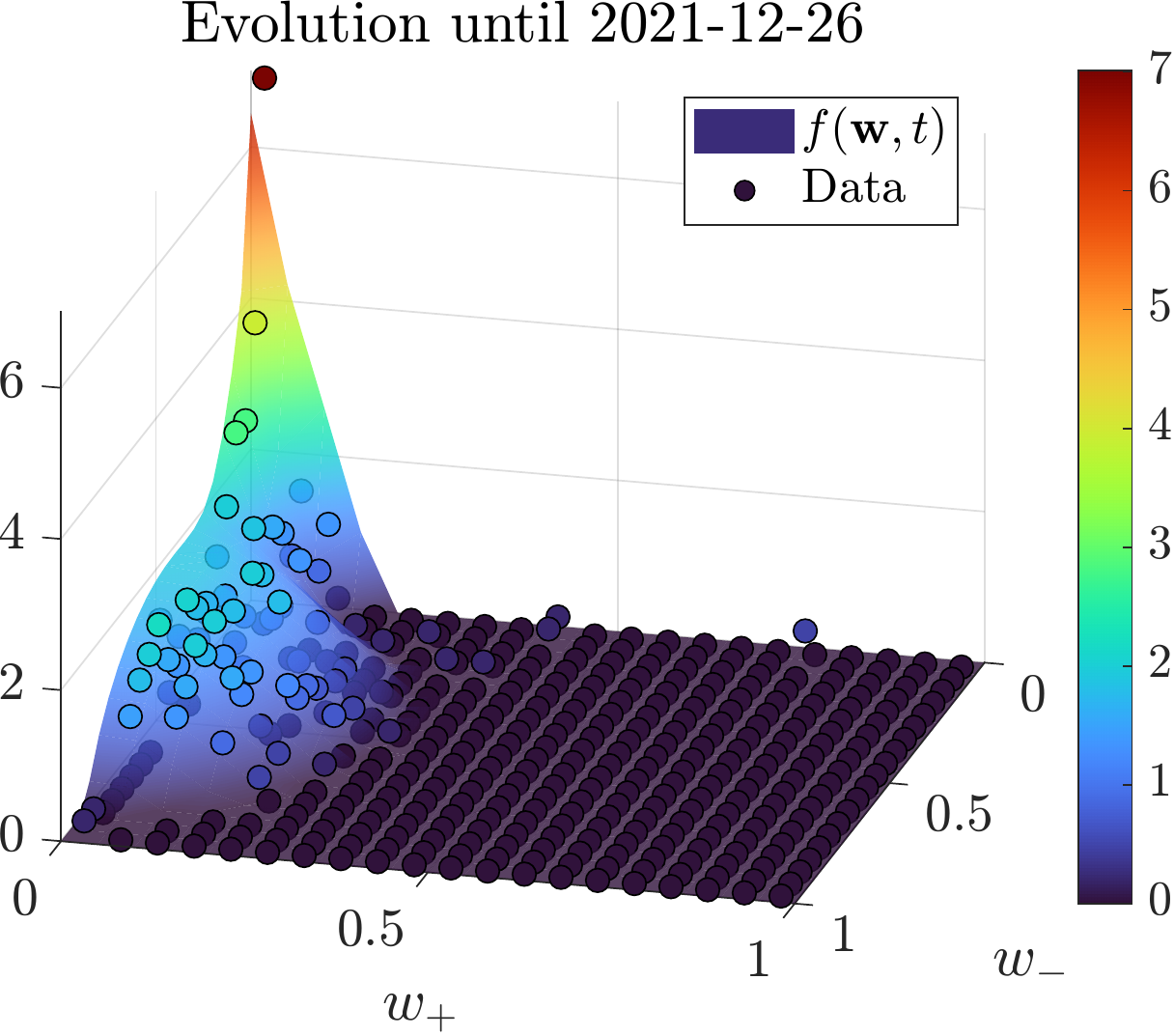}\hskip 0.5cm%
\includegraphics[width=0.4\linewidth]{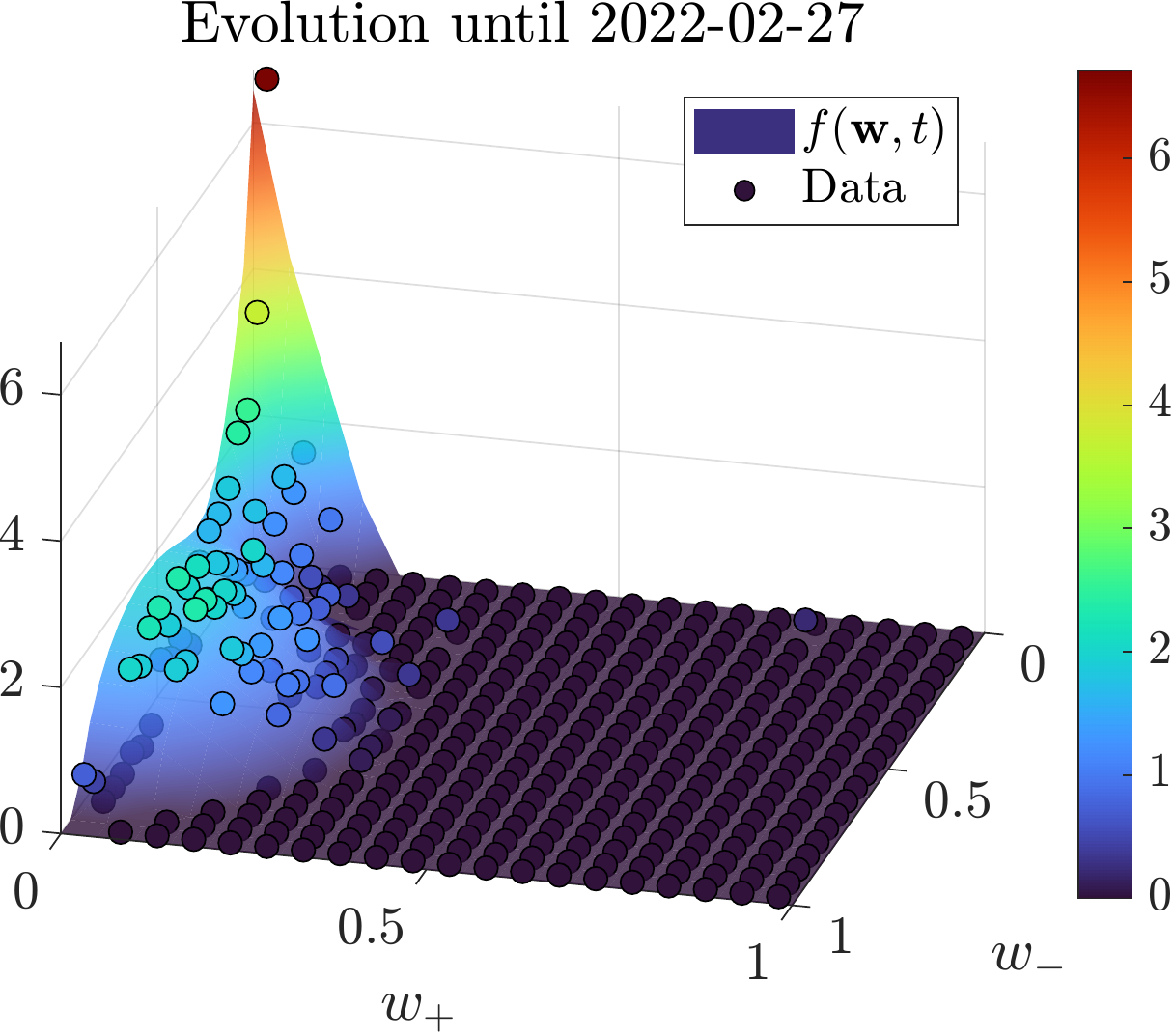}\hfil%
\caption{3D snapshots of dataset time series and model evolution. Again, base 
2-logarithm of values is used for coloring. The warmer colors in the bottom 
left quadrant show the increasing polarization effect around negative opinions.}
\label{fig:results3d}
\end{figure}

Each one-dimensional opinion direction was discretized using the scheme in^^>\cite{pareschi18} 
with a coarse grid of $20\times20$ points, whereas the time 
integration of the opinion evolution was computed with a semi-implicit scheme 
where the time step $\Delta t$ was chosen to be $O(\Delta\wpos)$, with 
$\Delta\wpos=\Delta\wneg$ being the steps for the 2D opinion domain, 
whereas the epidemic exchange portion was integrated through a simple explicit method.

\begin{figure}[t]
\centering
\includegraphics[width=0.4\linewidth]{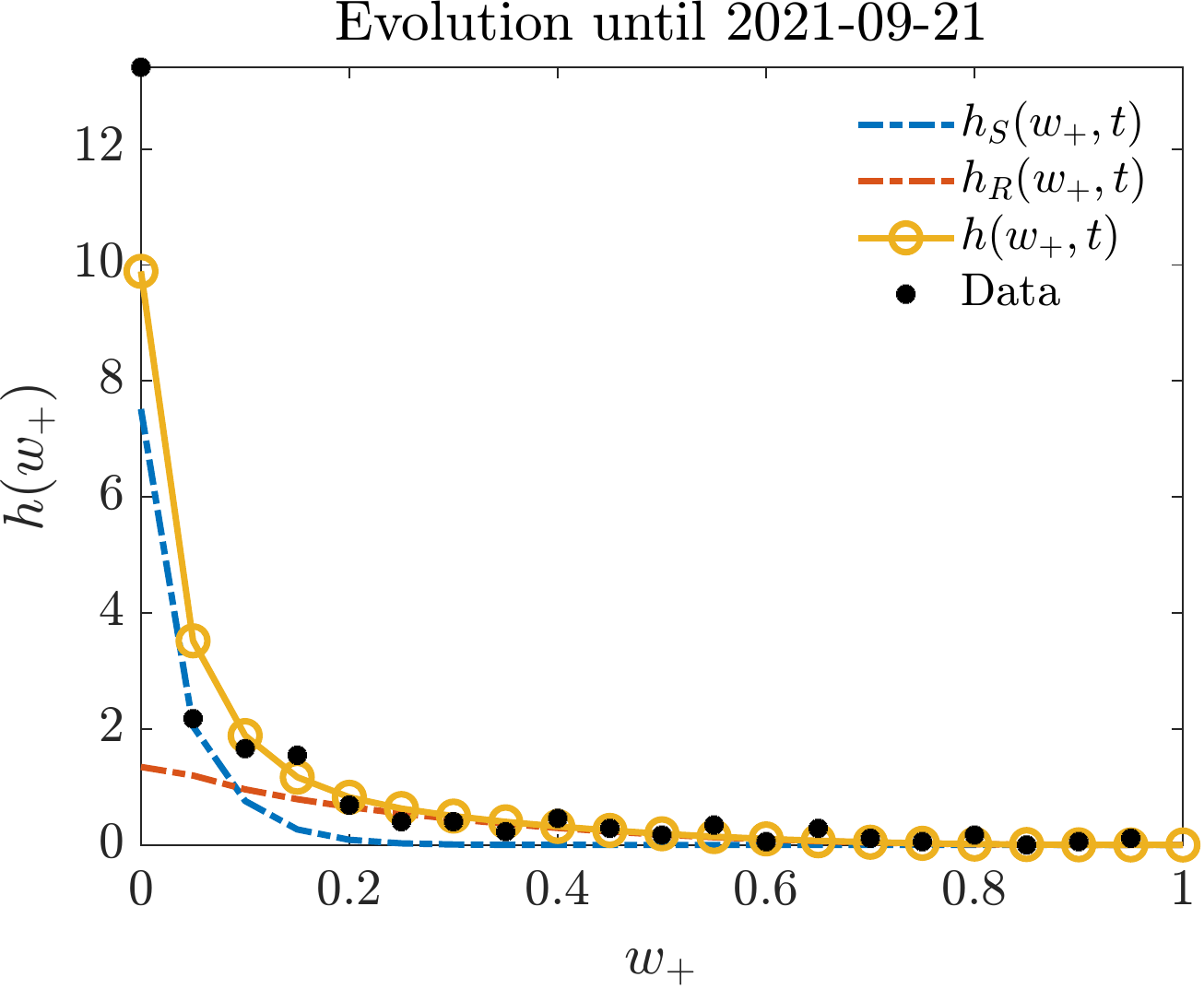}\hskip 0.5cm%
\includegraphics[width=0.4\linewidth]{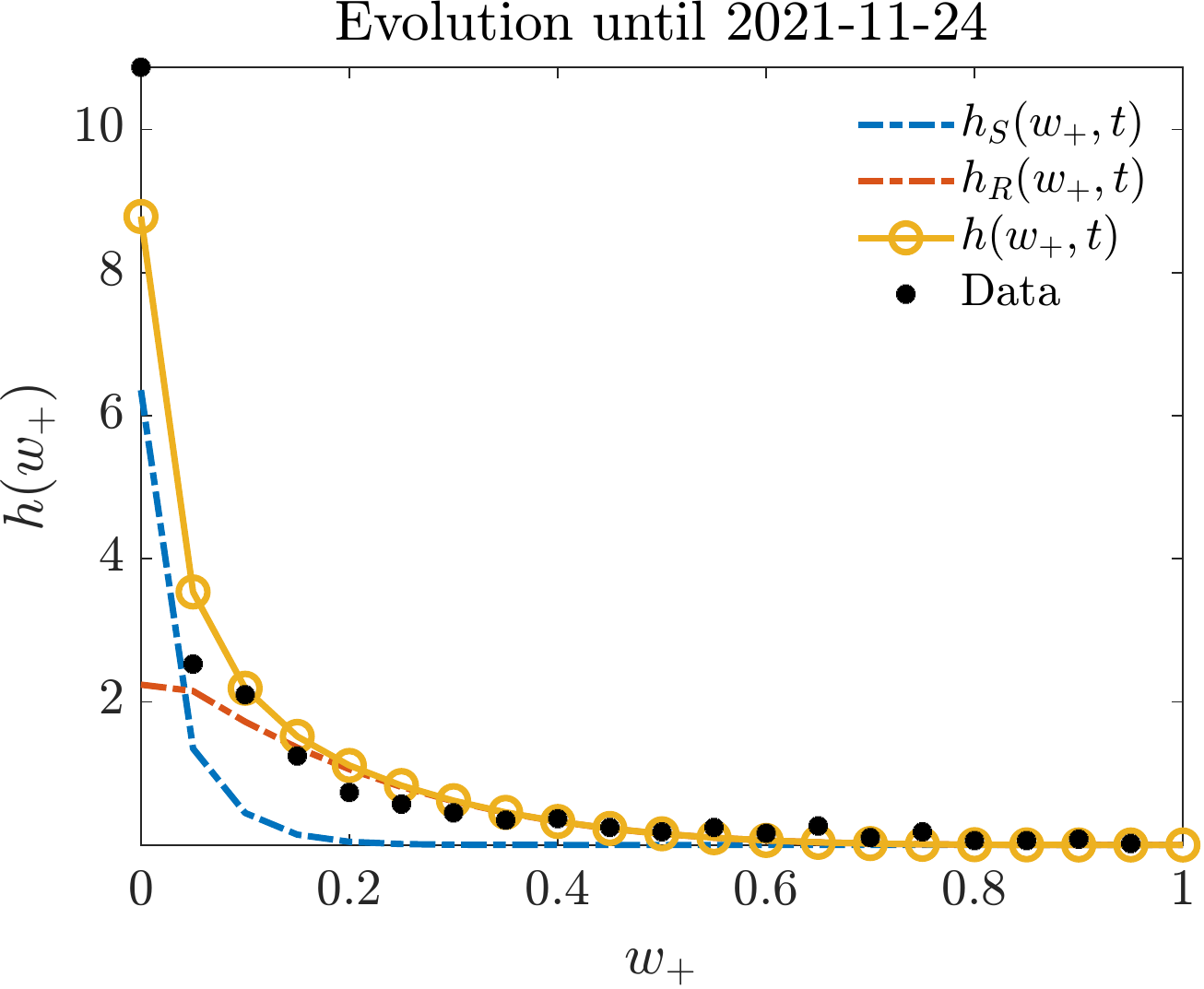}\\%
\includegraphics[width=0.4\linewidth]{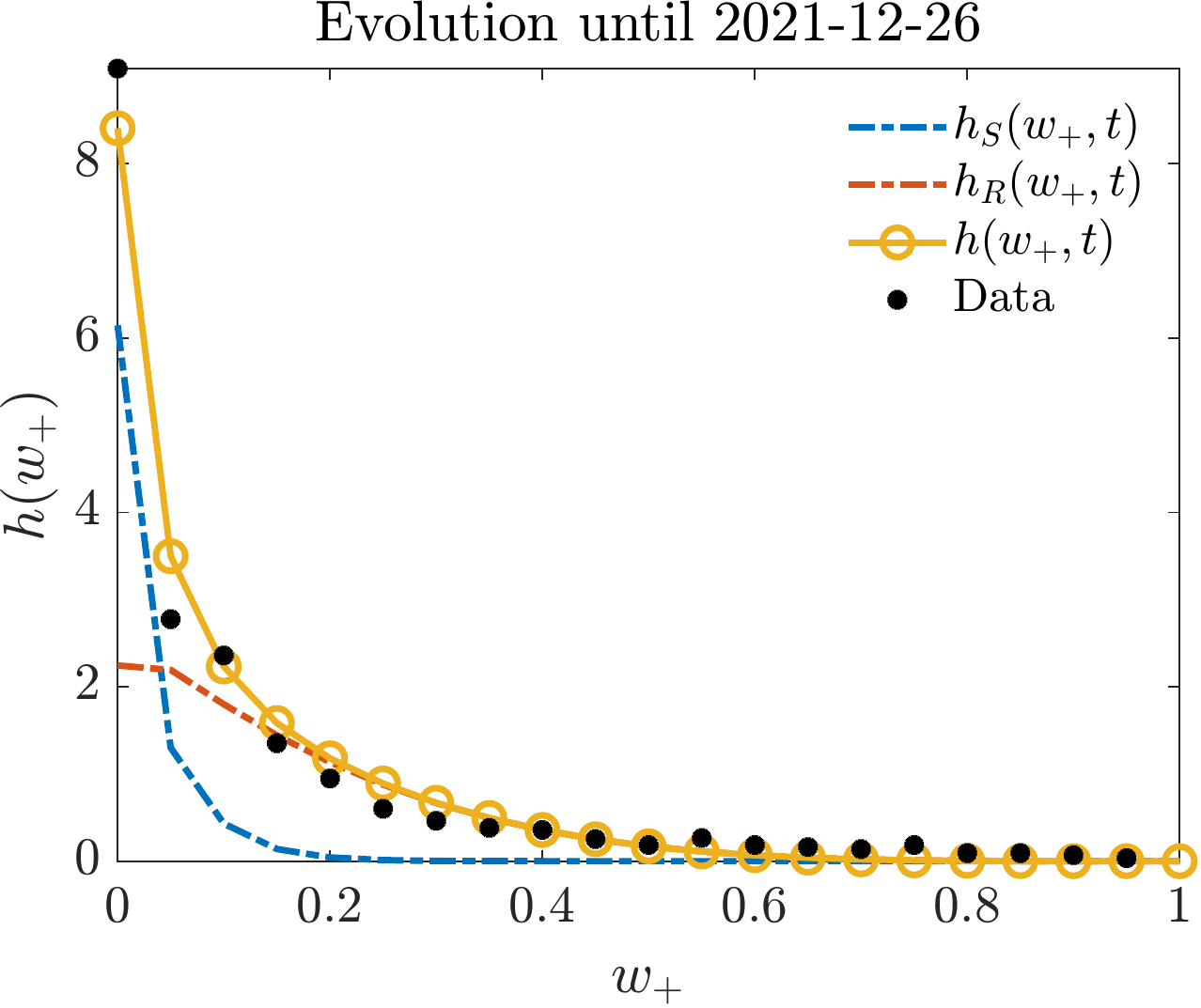}\hskip 0.5cm%
\includegraphics[width=0.4\linewidth]{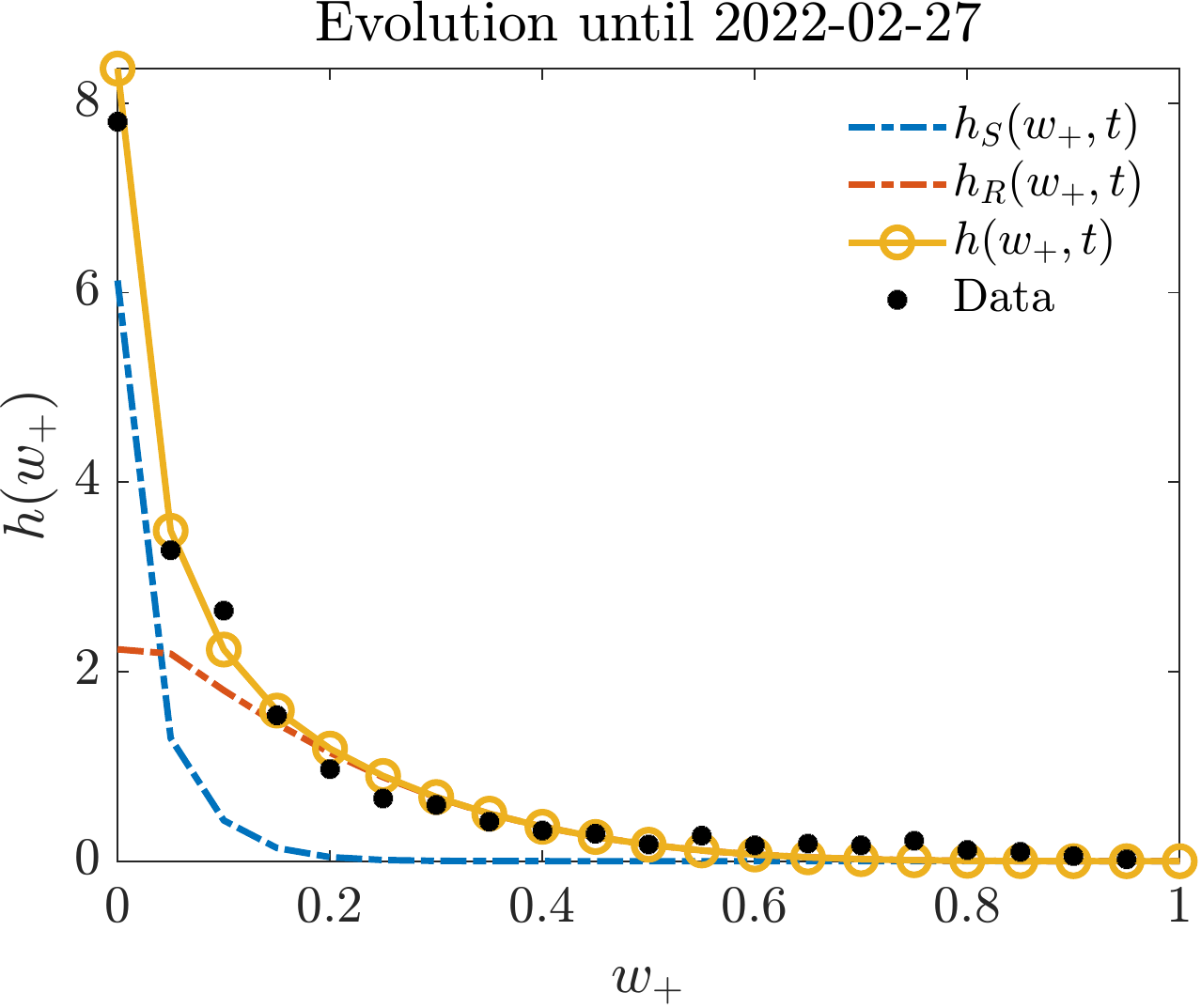}\hfil%
\caption{Evolution of the marginal density for the positive opinion: even if 
the total distribution appears to be unimodal, it is still the sum of two 
distinct profiles.}
\label{fig:resultmarginpos}
\end{figure}

\begin{figure}[t]
\centering
\includegraphics[width=0.4\linewidth]{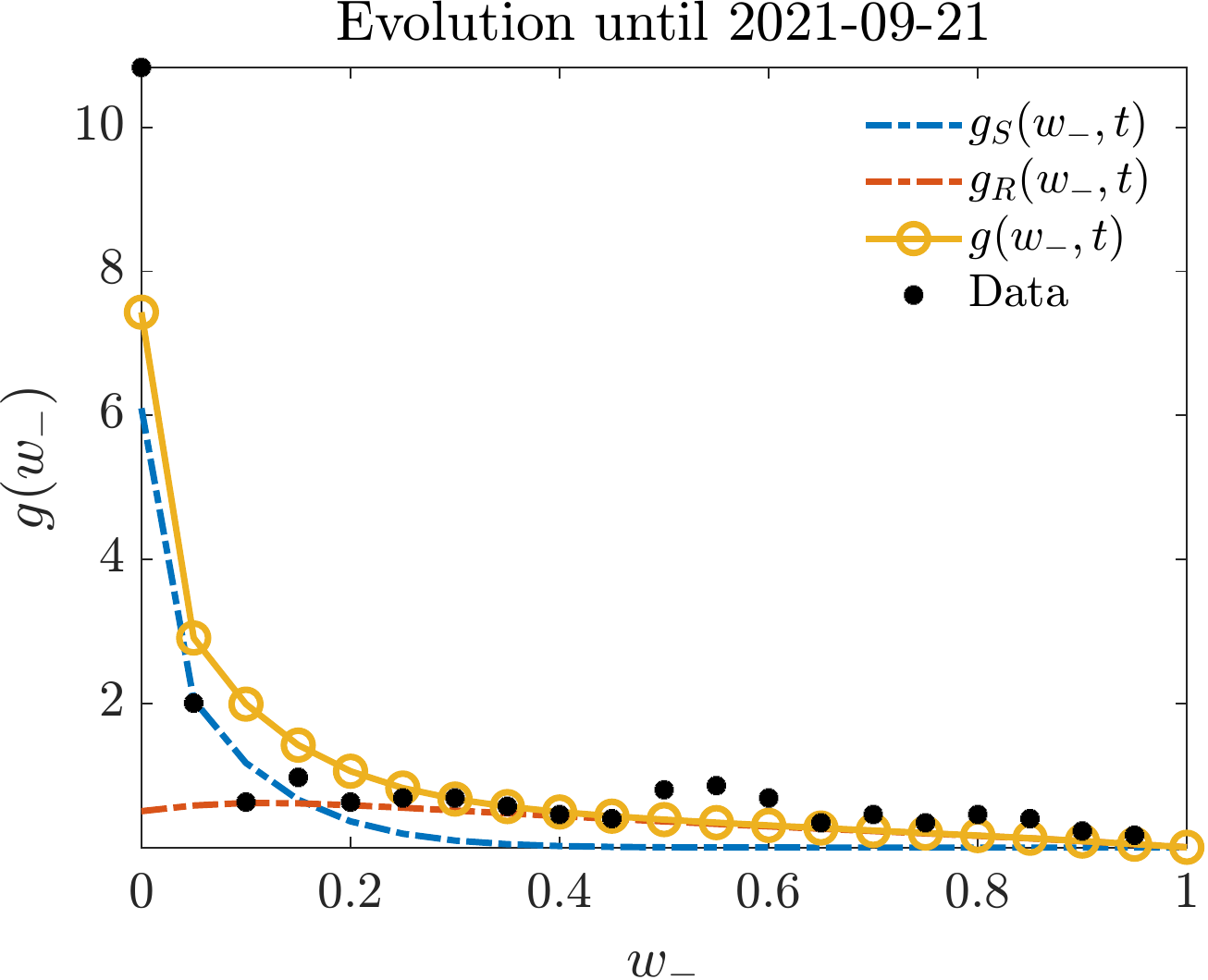}\hskip 0.5cm%
\includegraphics[width=0.4\linewidth]{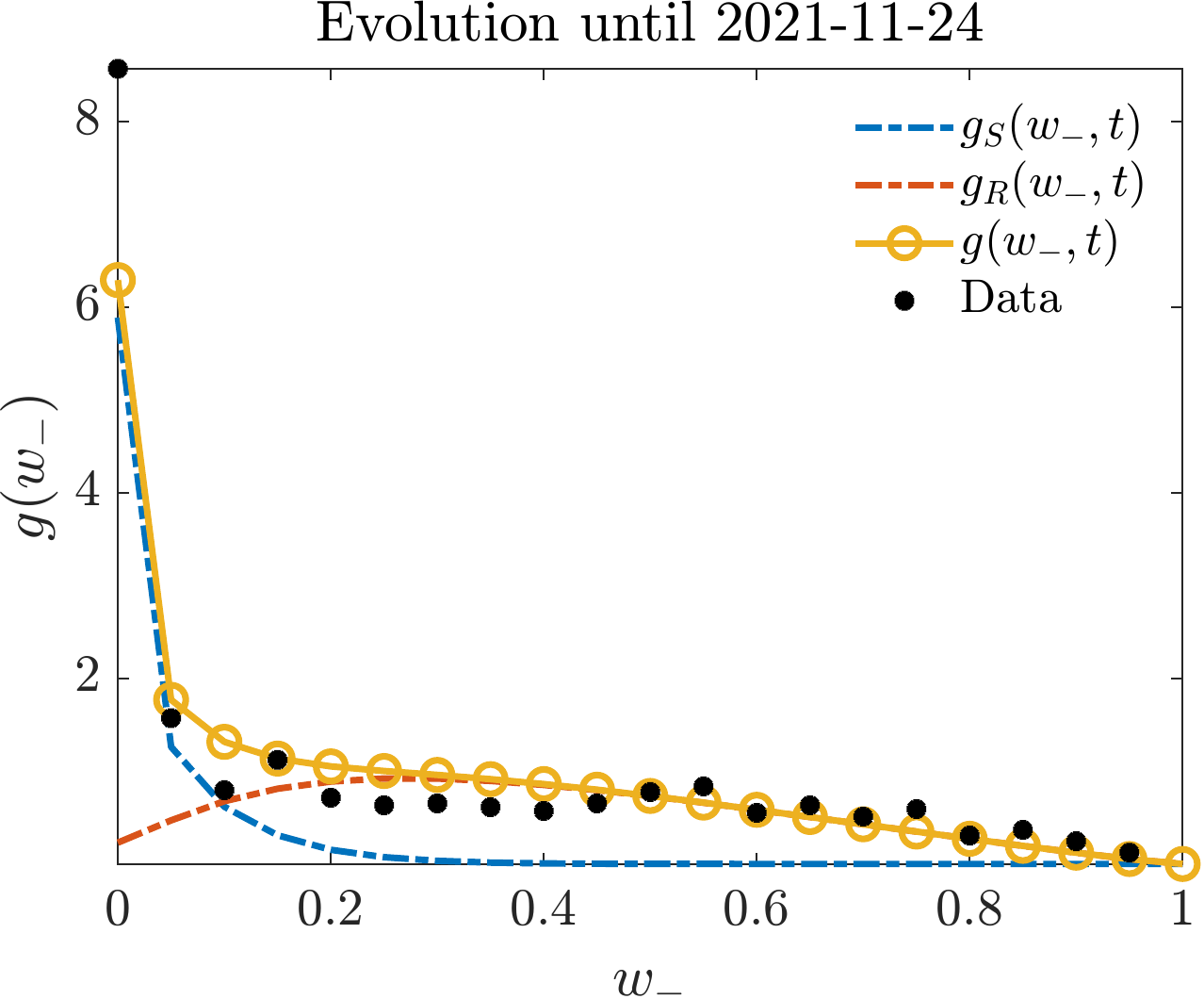}\\%
\includegraphics[width=0.4\linewidth]{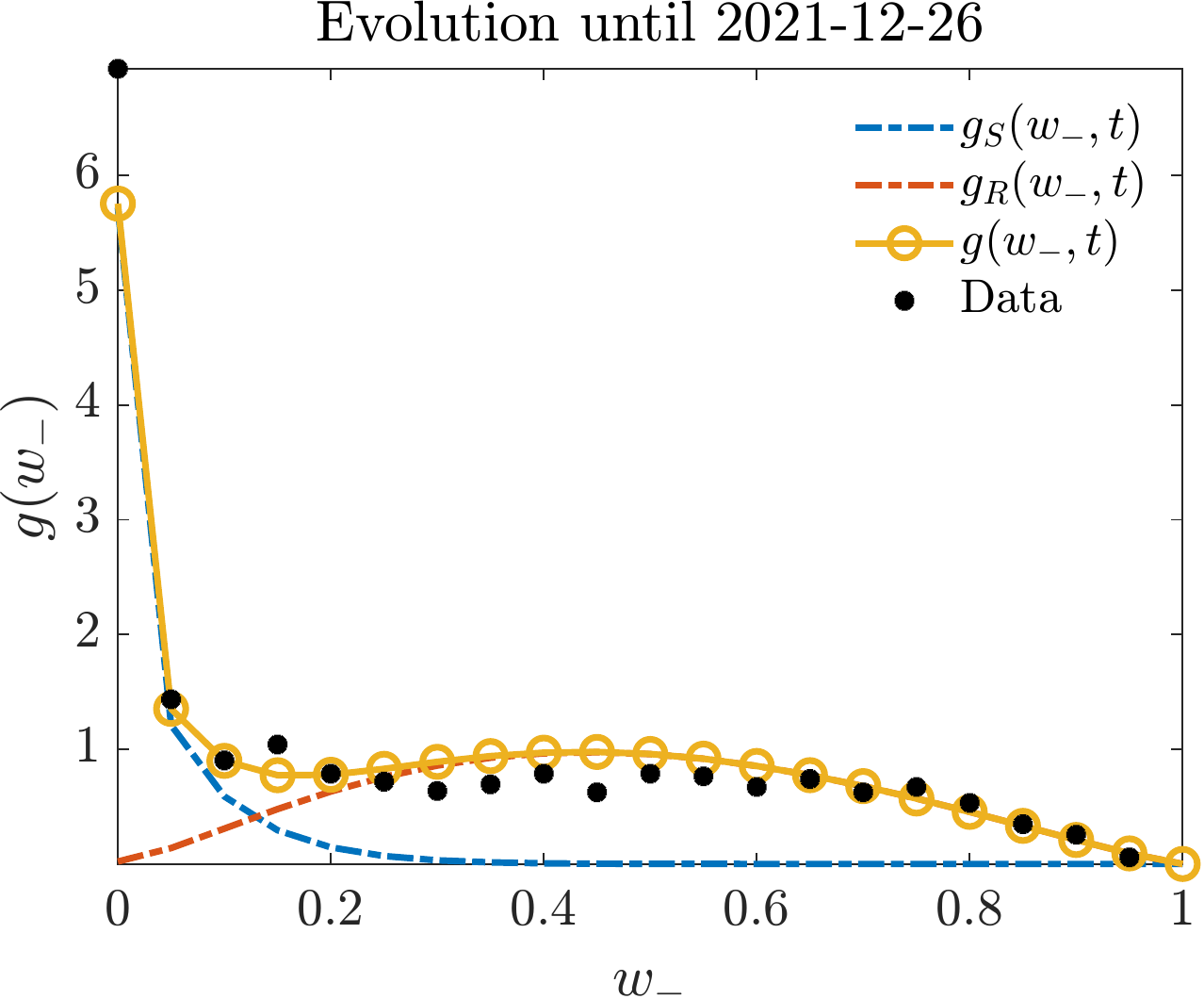}\hskip 0.5cm%
\includegraphics[width=0.4\linewidth]{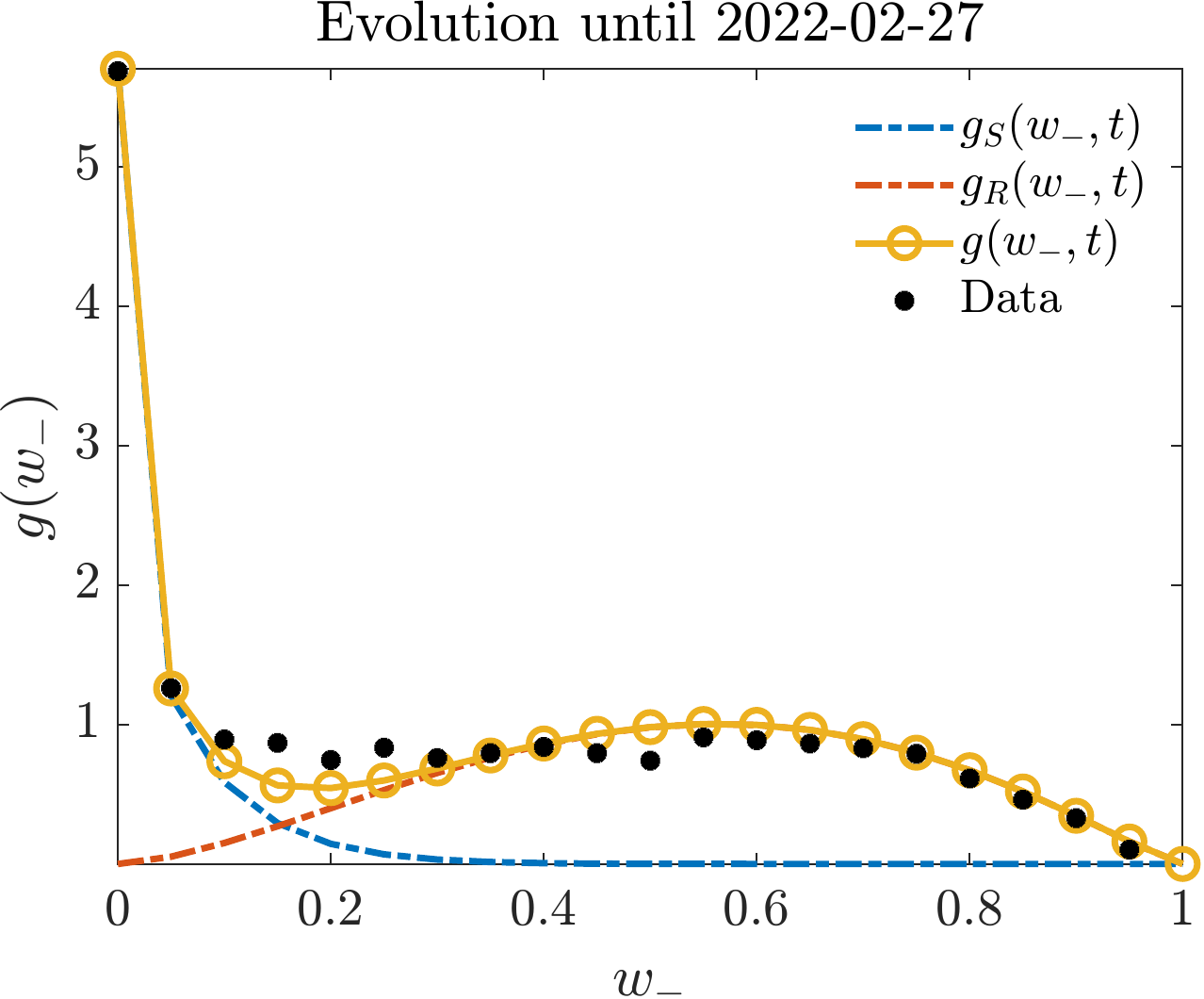}\hfil%
\caption{Evolution of the marginal density for the negative opinion: it is 
clear the emergence of a peak around the value of $w_\neg = 0.6$, i.e., a 
polarizing effect toward negative sentiments.}
\label{fig:resultmarginneg}
\end{figure}

In Fig.^^>\ref{fig:results3d} we report the comparison. Here, we can see that 
concentration around the origin starts very early, followed by a portion of 
data-points gathering towards a neighborhood of $(0.25,0.6)$ to form a peak 
later on. This polarization trend concerns only the negative opinions, as 
testified by the evolution of the marginal showed in 
Fig.^^>\ref{fig:resultmarginneg}; while the mixture of equilibria for the 
positive marginal keeps substantially the same profile of unimodal decrease 
(Fig.^^>\ref{fig:resultmarginpos}).

Overall, we can see that the model is capable of correctly identifying the 
formation and evolution of both unimodal and bimodal trends happening at the 
same time in the two-dimensional evolution. To sum up, the model can accurately 
predict the polarization process towards negative extreme shown in our dataset.

\section{Final considerations}
Mathematical modeling of fake-news spreading is a particularly timely and challenging topic, 
involving numerous areas of research with strong social impacts.
In this paper we focused on opinion formation processes within closed communities in presence of 
spreaders of misinformation. Inspired by a real case study from social data using NLP techniques, we 
presented a data-driven model based on vector stochastic differential equations. Then, in 
order to analyze the model and compute analytically the stationary solutions for its spatial 
marginals, we considered its mean-field approximation in the form of a system 
of Fokker-Planck equations, where the dissemination of fake news was carried on 
through a compartmental approach. Finally, we compared the evolution of the 
model computed numerically with the one of the dataset time series extract using
sentiment analysis. Our 
results show a good agreement between them, allowing us to observe the formation
of bimodal distributions indicating the polarization of opinions toward very 
negative sentiments as manifested in the real data. We emphasize that the present model, due to its generality, naturally lends itself to many other areas of application in relation to the analysis of fake-news dissemination using NLP techniques in different contexts.

\section*{Acknowledgments}
This work has been written within the activities of GNFM and GNCS groups of INdAM (National Institute of High Mathematics). 
JF acknowledges partial support of MUR-PRIN2020 Project No. 2020JLWP23 
\lq\lq Integrated mathematical approaches to socio-epidemiological 
dynamics\rq\rq. All authors acknowledge the support
of the FIR 2021 project \lq\lq No hesitation. For an effective communication of 
the Covid-19 vaccination\rq\rq.

%

\end{document}